\documentclass[aps,prb,twocolumn,showkeys,superscriptaddress,groupedaddress,nofootinbib,floatfix]{revtex4}
\usepackage{inputenc}
\usepackage{natbib}
\usepackage{tabularx}
\usepackage{graphicx}  
\usepackage{epsfig}
\usepackage{float}
\usepackage{dcolumn}   
\usepackage{array}
\usepackage{bm}        
\usepackage{amssymb}   
\usepackage{amsmath}
\usepackage{amstext}
\usepackage{mathtools}
\usepackage{footnote}
\usepackage{threeparttable}
\usepackage{siunitx} 
\usepackage{multirow}
\usepackage{lipsum}
\usepackage{mathrsfs}
\usepackage{physics}
\usepackage{tikz} 
\usepackage{textcomp}
\usepackage{xcolor, soul}
\soulregister\cite7
\soulregister\ref7
\soulregister\pageref7
\usetikzlibrary{shapes,fit,backgrounds}
\usetikzlibrary{positioning}
\usetikzlibrary{shadows,fadings}
\usetikzlibrary{shapes.callouts} 
\usetikzlibrary{arrows}

\usepackage [colorlinks=true,allcolors=blue]{hyperref}

\DeclareRobustCommand{\element}[1]{\@element#1\@nil}
\def\@element#1#2\@nil{%
  #1%
  \if\relax#2\relax\else\MakeLowercase{#2}\fi}
\makeatother

\begin{document}
\title{Electron-lattice interplays in LaMnO$_3$ from canonical Jahn-Teller distortion notations}
\author{Michael Marcus Schmitt}
\email []{MMN.Schmitt@doct.uliege.be}
\affiliation{Physique Theorique des Materiaux, Q-Mat, CESAM, Universite de Liege, Allee du 6 Aout 17 (B5), 4000 Sart Tilman, Belgium}
\author{Yajun Zhang}
\affiliation{Physique Theorique des Materiaux, Q-Mat, CESAM, Universite de Liege, Allee du 6 Aout 17 (B5), 4000 Sart Tilman, Belgium}
\affiliation{Department of Engineering Mechanics, School of Aeronautics and Astronautics, Zhejiang University, 38 Zheda Road, Hangzhou 310007, China}
\author{Alain Mercy} 
\affiliation{Physique Theorique des Materiaux, Q-Mat, CESAM, Universite de Liege, Allee du 6 Aout 17 (B5), 4000 Sart Tilman, Belgium}
\author{Philippe Ghosez}
\affiliation{Physique Theorique des Materiaux, Q-Mat, CESAM, Universite de Liege, Allee du 6 Aout 17 (B5), 4000 Sart Tilman, Belgium}
\date{\today}
\begin{abstract}
LaMnO$_3$ is considered as a prototypical Jahn-Teller perovskite compound, exhibiting a metal to insulator transition at $T_{JT} = 750K$ related to the joint appearance of an electronic orbital ordering and a large lattice Jahn-Teller distortion. From first-principles, we revisit the behavior of LaMnO$_3$ and show that it is not only prone to orbital ordering but also to charge ordering.  Both charge and orbital orderings appear to be enabled by rotations of the oxygen octahedra and the subtle competition between them is monitored by a large tetragonal compressive strain, that is itself a Jahn-Teller active distortion. Equally, the competition of ferromagnetic and antiferromagnetic orders is slave of the same tetragonal strain. Our results further indicate that the metal to insulator transition can be thought as a Peierls transition that is enabled by spin symmetry breaking. Therefore, dynamical spin fluctuations in the paramagnetic state stabilize the insulating phase by the instantaneous symmetry breaking they produce and which is properly captured from static DFT calculations. As a basis to our discussion, we introduce canonical notations for lattice distortions in perovskites that distort the oxygen octhedra and are connected to charge and orbital orderings.
\end{abstract}
\keywords{LaMnO$_3$, perovskites, Jahn-Teller distortions, first-principles calculations}
\maketitle

\section{Introduction}

Since the discovery of the colossal magnetoresistance effect in R$^{3+}_x$A$^{2+}_{1-x}$MnO$_3$ manganese perovskites solid solutions  about 25 years ago\cite{Helmolt1993} there has been a continuous research effort to understand the physical behavior of the end-members as well as intermediate compounds. Nonetheless, for the rare earth manganite perovskite side, RMnO$_3$, no fully consistent picture has emerged yet that explains the interplay between structural, magnetic, and electronic degrees of a freedom. Hence, the prototypical member of this series, LaMnO$_3$, still attracts an extensive research interest. 

LaMnO$_3$ belongs to a large class of perovskite materials with a Goldschmidt \cite{Goldschmidt1926} tolerance factor  $\textit{t} < 1$. As such its lattice structure deviates from the ideal cubic perovskite $Pm\overline{3}m$ reference phase by the appearance of cooperative rotations of the MnO$_6$ oxygen octahedra.
Above 1200K, LaMnO$_3$ shows a rhombohedral space group $R\overline{3}c$\cite{3539fa633bee47ac9231ce02c2ca1b41,Qiu2005}, with rotations of the connected oxygen octahedra according to a $a^-a^-a^-$ rotation pattern (in Glazer's notations \cite{Glazer1972}).
At 1200 K, LaMnO$_3$ undergoes a structural phase transition to a \textit{Pbnm} phase with $a^-a^-c^+$ rotation pattern, the most common one among the perovskites\cite{Lufaso2001}.

In both of these phases, oxygen octahedra rotate in a nearly rigid way. This rigid rotation preserves the cubic symmetry ($O_h$ in Sch\"onflies notation) around the Mn atom if only the octahedron is considered. 
In such a regular octahedron, the five-fold degenerated Mn \textit{d-} states are split into three degenerate lower energy $t_{2g}$ and two degenerate higher energy $e_g$ states.  In the 3+ oxidation state of Mn, four electrons formally occupy the Mn-\textit{d} states. Due to strong intra-site Hund's coupling in th $3d$ shell, Mn adopts a high-spin configuration, where three electrons occupy the $t_{2g}$ and one the $e_g$ states.
As the Mn-$3d$ states constitute the highest occupied states in LaMnO$_3$, it is consequently metallic in the  $R\overline{3}c$ and $Pbnm$ phases at high temperatures.

At 750K and ambient pressure, or lower temperatures and higher pressure ($\approx 32\si{\GPa}$), a second structural transition occurs, accompanied by a metal-to-insulator transition (MIT). This transition is called Jahn-Teller (JT) or Orbital Ordering (OO) transition at the temperature $T_{JT}$ or $T_{OO}$\cite{Baldini2011}. At this transition, a sudden increase of volume is observed. The initially nearly cubic unit cell shows a strong tetragonal compression and orthorhombic deformation\cite{ Chatterji2003, Maitra2004,Ahmed2009}. The oxygen octahedra experience strong cooperative deformations lowering their symmetry from cubic to orthorhombic ($O_h$ to $D_{2h}$). These are the so-called \textit{Jahn-Teller distortions}. However, no further symmetry reduction occurs and the structure preserves the \textit{Pbnm} space group\cite{Rodriguez-Carvajal1998}.
Hence, the structures are called $O'$($T<T_{JT}$) and $O$ ($T>T_{JT}$)\cite{Sanchez2003,Qiu2005}. A peculiarity of such isosymmetrical transitions is that the structural order parameter - the Jahn-Teller distortions - are not restricted to zero amplitude before the transition. Consequently in the $O$ phase local  Jahn-Teller distortions are reported and short-range ordered clusters with the diameter of 4 MnO$_6$ octahedra have been found \cite{Sanchez2003,Qiu2005,Thygesen2017}.
 
In all of the above described phases, the unpaired magnetic moments in the $3d$ shell of manganese are disordered and LaMnO$_3$ is paramagnetic (PM). At $T_N=140K$\cite{Moussa1996}, LaMnO$_3$ undergoes a magnetic transition without any structural changes to an antiferromagntic phase with A-type pattern (AFM-A). 

There is a long standing debate about the origin of the MIT at $T_{JT}$ in LaMnO$_3$\cite{Kovaleva2004,Yamasaki2006,Kovaleva2010,Baldini2011,Nucara2011,Sherafati2016}. Broadly, this debate can be summarized into two distinct views: the \textit{cooperative Jahn-Teller Effect}\cite{Englman1970,Halperin1971,englman1972jahn,Gehring1975} (C-JTE) on the one hand and the spontaneous orbital ordering proposed by the \textit{Kugel-Khomskii}\cite{ki1982jahn} (KK) model on the other hand. 

The C-JTE approach extents the Jahn-Teller Effect\cite{Jahn1937} from an isolated Jahn-Teller center to a solid of coupled centers. In the case of LaMnO$_3$, these are the corner shared oxygen octahedra. The origin of the transition is the \textit{local} degeneracy of the $e_g$ orbitals, which induces a \textit{local} octahedral distortion removing the degeneracy. The coupled octahedra only interact harmonically through their individual deformation. The cooperative ordering of the octahedra results of the minimization of the lattice harmonic energy and creates an orbital ordering. 

The KK approach (based on the Mott-Hubbard model\cite{Hubbard1963}) emphasizes instead the roles of the \textit{inter site} super-exchange electronic interactions and dynamical correlations between $e_g$ electrons. It postulates spontaneous orbital and magnetic orderings in the undistorted cubic perovskite phase for a certain ratio of hopping and exchange parameters. The appearance of the cooperative deformation of the oxygen octahedra is here a secondary effect induced by the orbital ordering. From DFT+DMFT calculations, it has been shown that the KK mechanism alone cannot account for the orbital ordering in LaMnO$_3$ \cite{Pavarini2010} and that electron-lattice coupling is crucial in promoting the high orbital ordering transition-temperature. Moreover a recent first-principles study\cite{varignon2019origin} claims that dynamical correlations are not necessary to account for orbital ordering in perovskites. LaMnO$_3$ thereby appears as a special case, where the principal orthorhombic Jahn-Teller distortion is only unstable in the presence of octahedral rotations. 

In the present work, we reinvestigate LaMnO$_3$ from  first-principles calculations. First, we show that our calculation method properly reproduces a range of measured ground-state properties of LaMnO$_3$. Then, we sample the Born-Oppenheimer potential energy surfaces (PES) of the close competing AFM-A and ferromagnetic (FM) orders and characterize the inherent electronic instabilities, couplings between phonon modes, strains, insulating and metallic states. By a simple Monte-Carlo (MC) simulation we show that these PESs qualitatively reproduce the orbital ordering transition at 750K. Finally, we unveil that LaMnO$_3$ shows an inherent subtle competition between charge-ordering and orbital ordering, which was suspected before\cite{Moskvin2009}. As a support to our analysis, we reclassify \textit{all} octahedra deforming cooperative distortions in perovskite systems into unified \textit{canonical} notations for those kind of distortions taking into account local and global aspects and show the connection to other various notations in the present literature.

The analysis of the PESs computed from DFT shows that the large electron-lattice coupling, necessary to explain the transition, lies in a Peierls effect \cite{peierls1991more}. The large coupling is only enabled once the spin symmetry between neighbouring sites is broken. At elevated temperature this symmetry breaking is produced by dynamical fluctuations of orbital occupations and spin orientations in the PM state, so questioning the ability of DFT to describe such a phenomenon. From MC simulation relying on PESs calculated with DFT and reproducing the MIT, we show that, in line with Ref. \cite{Varignon2019}, the key ingredient of the MIT transition is more the {\it instantaneous} symmetry breaking, that can be statically treated in DFT, than the {\it dynamical} nature of the fluctuations. Hence, we assign an important part of the stabilising energy at the MIT transition to \textit{spin symmetry breaking}. This does not mean that dynamical electron correlations do not play any role in stabilising the insulating phase. The use of an appropriate U-correction remains important in treating materials like LaMnO$_3$ from DFT. Nevertheless, together with the recent explanation of charge-ordering in $e_g^1$ alkaline earth ferrites AFeO$_3$ \cite{Zhang2018} and rare earth RNiO$_3$ \cite{Mercy2017} as a Peierls transition, it seems that the cooperative Jahn-Teller/orbital ordering and charge-ordering transitions are simply different coordinates for translational symmetry breaking and therefore they might always compete in perovskites with degenerate $e_g$ states. Finally, we emphasize that for gaining more insights in the dynamical properties of the MIT, new model descriptions are needed that can treat the electrons and nuclei dynamically coupled in large supercells.


\section{Methods}
\label{sec:III_Methods} 

Density functional theory (DFT) calculations were performed using the generalized gradient approximation (GGA) with the revised Perdew-Burke-Enzerhof parameterisation for solids (PBEsol) \cite{Perdew2008}  as implemented in the Vienna ab initio simulation package (VASP) \cite{Kresse1999}. A Liechtenstein $(U|J)$ correction was applied. $(U|J) = (5|1.5)$ were determined by comparing structural, electronic, and magnetic parameters to experimental results. For comparison, we reproduced also the results of \textit{Mellan et al.} using $(U|J) = (8|2)$ \cite{Mellan2015}. The projector augmented wave method\cite{Bloechl1994}  was used, with a high plane-wave cutoff energy of 600 \si{\eV} and a dense 14x14x14  Monkhorst-Pack k-point mesh\cite{PhysRevB.13.5188} with respect to the cubic perovskite unit cell. Supercells up to 40-atoms were used to treat various magnetic orderings. The density of the k-point mesh in the supercells was reduced according to the multiplicity of the supercell. During the structural optimizations, the lattice parameters and internal coordinates of atoms were fully relaxed until the Hellmann-Feynman forces on each atom were less than $10^{-5} \si{\eV/\angstrom}$ and stresses are less than $4\cdot10^{-4}$ \si{\electronvolt/\angstrom^3}.

We used \textsc{isodistort}\cite{Campbell2006} to analyze symmetry-adapted modes and symmetry-adapted strains of experimental and optimized structures. In all cases, we used the aristotype $Pm\overline{3}m$ structure of LaMnO$_3$ as reference, with a lattice constant of $a_0=3.935\si{\angstrom}$ that preserves the same volume per formula unit as in the experimental \textit{Pbnm} phase at low temperatures. Then, we used the software \textsc{invariants}\cite{Hatch2003} to create invariant coupling terms including symmetry adapted modes and strains. We use the \textsc{bandup} utility\cite{Medeiros2014,Medeiros2015} to unfold electronic band-structures of magnetically or structurally distorted structures back to the Brillouin-zone of the cubic 5-atoms perovskite unit-cell. Finally we used an in-house tool to approximate PESs from DFT data with a polynomial expansion and to run Monte-Carlo simulations on the determined polynomial.  


\section{Canonical notations for cooperative Jahn-Teller Distortions in perovskites}
\label{sec:II_Redef_JTD}
The Jahn-Teller effect in the ideal perovskite $Pm\overline{3}m$ space group has been intensively studied over decades. Surprisingly, no unified notation of cooperative Jahn-Teller distortions has been adopted yet. The reason for that seems to be the focus of many works on limited subsets of distortions, for which labels are defined in the scope of the work. Here, we introduce canonical notations defining a \textit{unique} label for \emph{all} possible distortions. These are beyond the scope of the investigated problems in LaMnO$_3$, but will serve to simplify future discussions and comparisons between different perovskites. The new labels combine local and cooperative aspects, while being based on existing notations. As a starting point we give a brief summary on the history of the study of the Jahn-Teller effect in octahedral transition-metal complexes.

In 1937 \textit{Jahn} and \textit{Teller} published a pionnering work stating that in a molecule "\textit{stability and} (orbital) \textit{degeneracy are not possible simultaneously unless the molecule is a linear one} [...]." \cite{Jahn1937}. The geometric instability of a molecule containing an orbital degenerate state is introduced by the so-called vibronic-coupling terms. These terms couple the degenerate electronic state linearly to a vibrational mode coordinate $Q_k$. The strength of the coupling is expressed as 

 \begin{equation}
 \alpha_{JT} = \mel**{\Psi^0_i}{\frac{\partial H_0}{\partial Q_k}}{\Psi^0_j},
 \label{eq:Vib_coupl}
 \end{equation}
 where $\Psi^0_i$, $\Psi^0_j$ are degenerate electronic states in a high-symmetry structure of the molecular system and $H_0$ is the Hamiltonian of the unperturbed system.  

Shortly after, the combinations of orbitals and modes that fulfill the symmetric conditions for such an effect in specific point groups were determined. \textit{Van Vleck}\cite{Vleck1939} studied the isolated octahedral transition-metal complex $MX_6$ (Point Group $O_h$) within an external crystal field. From the 21 normal modes (3 times 6 atomic displacements plus 3 rigid rotations of the oxygen octahedron with respect to the external field), he identified six that are prone to a Jahn-Teller instability in conjunction with degenerate $t_{2g}$ and/or $e_g$ orbitals and labeled them from $Q_1$ to $Q_6$: $Q_1$, the volume expansion/contraction, $Q_2$ a planar rhombic distortion, $Q_3$ the tetragonal distortion, where $Q_2$ and $Q_3$ keep the octahedral volume constant at linear order, and $Q_4$ to $Q_6$ the three possible shears of the octahedron (see Table \ref{Tab:Define_Qi}). 

At the molecular level, $Q_1$ does not play any role if the reference volume of $O_h$ point group represents a stationary point with respect to volume expansion/contraction. Moreover, it does not lift the electronic degeneracy as it keeps the symmetry of the $O_h$ group.

The modes $Q_2$ and $Q_3$ are degenerate and posses the $E_g$ symmetry with respect to $O_h$. In conjunction with the $e_g$ orbitals ($dz^2-r^2, dx^2-y^2 $), they form the extensively studied  $E_g \otimes e_g$ Jahn-Teller system. Large static $Q_2/Q_3$ distortions appear for oddly occupied $e_g$ orbitals as e.g. Mn$^{3+}(e_g^1)$ or Cu$^{2+}(e_g^3)$. At the harmonic level, the systems forms the so called \emph{mexican hat} potential energy surface. This surface possesses a degenerate minimum described by a circle in the $Q_2-Q_3$ plane. Which point on the circle is stabilized depends then on the strength and sign of higher order anharmonicities \cite{Oepik1957,o1993jahn,Garcia-Fernandez2005}. The amplitudes of the distortion are quantified by 
\begin{align} 
Q_2 = \frac{2(l-s)}{\sqrt{2}} \\ 
Q_3 = \frac{2(2m-l-s)}{\sqrt{6}}
\end{align}  
where $l$, $m$, and $s$, refer to long, middle, and short $MX$ bond lengths.
The angle in the $Q_2/Q_3$ plane is 
\begin{equation}
	\phi = \arctan(\frac{Q_2}{Q_3})
\end{equation}
and is a direct measure for the $dz^2-r^2/dx^2-y^2$ ratio in the stabilized state.
 
The modes $Q_4$ to $Q_6$ are relevant for degenerate $t_{2g}$ states, since they posses the same symmetry and form a $T_{2g} \otimes t_{2g}$ system. However, the $t_{2g}$ orbitals can also interact with $E_g$ modes ($Q_2$ and $Q_3$), which results in many possibilities for energy lowering distortions to a degenerate $t_{2g}$ system. Spin-orbit coupling further complicates the situation for heavier transition-metal ions as it introduces a splitting opposed to the distortion  \cite{Sturge1968,khomskii2014transition}. The vibronic couplings are typically small since the strength of $\pi$-bonds formed between the $M$ $t_{2g}$ orbitals and neighboring $X$ $p$-orbitals are weak. Consequently, smaller static distortions typically appear in systems with degenerate $t_{2g}$ states than in those with degenerate $e_g$ states.

The problem of the Jahn-Teller instability in isolated $MX_6$ octahedra was soon transferred to periodic solids, in which each unit cell contains a Jahn-Teller ion. Amongst them are the $ABX_3$ perovskites with their corner-shared BX$_6$ octahedral network. Jahn-Teller instabilities occur in $ABX_3$ perovskites with an odd occupation of the B-cation's $e_g$ orbitals, such as rare-earth manganites RMnO$_3$ ($d^4 = e_g^1$), KCrF$_3$ ($d^4 = e_g^1$)\cite{Margadonna2006}, KCuF$_3$ ($d^9 = e_g^3$)\cite{Lufaso2004}, or with an incomplete occupation of the $t_{2g}$ orbitals such as rare-earth titanates RTiO$_3$ ($d^1=t_{2g}^1$)\cite{Komarek2007,Varignon2017} and rare-earth vanadates RVO$_3$ ($d^2 = t_{2g}^2$)\cite{Carpenter2009a}. The key difference between the isolated problem studied by \textit{Van Vleck} and the perovskites with connected Jahn-Teller centers lies in the direct neighboring of the Jahn-Teller ions. As a first consequence, the lattice of sites implies that the degenerate electronic states form continuous electronic bands instead of well-defined orbital states. The electronic band character of the degenerate states has been largely ignored by the C-JTE and KK theories. The C-JTE approach directly transfers the Jahn-Teller Hamiltonian of the isolated problem to the periodic solid by simply exchanging the normal modes with phonon-type modes and lattice strains \cite{Kanamori1960,Halperin1971,englman1972jahn,Gehring1975}. In the KK view the band-character is quasi ignored by an assumption of very small bandwidths\cite{ki1982jahn}. As a second consequence, individual distortions are transferred between octahedral sites. The network allows nonetheless for some phase freedom in the cooperative arrangement of the distorted octahedra. This additional freedom enables the system to achieve the same individual octahedral distortion from different cooperative orderings.   

\begin{table*}
    \renewcommand{\arraystretch}{1.15}
    \centering 
	\caption{Canonical labels $Q_{i\alpha}^{\vec{q}}$ for cooperative Jahn-Teller distortions in solids with octahedral corner shared networks. The first subscript $i$ refers to the \textit{Van Vleck's} numbering of normal modes in the isolated octahedron. The second subscript $\alpha$ defines the unique axis of the local distortion pattern. $\alpha$ is not necessary for the isotropic deformations $Q_1^{\mathbf{\Gamma}}$ and $Q_1^{\mathbf{R}}$. The superscript $\vec{q}$ refers to the reciprocal space vector with which the mode is translating. Shown are $\mathbf{\Gamma}=(0,0,0)$, $\mathbf{X}=($\textonehalf$,0,0),\mathbf{M}=($\textonehalf,\textonehalf$,0)$, and $\mathbf{R} = ($\textonehalf,\textonehalf,\textonehalf$)$. $\mathbf{\Gamma}$ is associated to lattice strains. Octahedra drawn in red and blue experience opposite distortions.}
    \label{Tab:Define_Qi}
    \vspace{0.5em}
    \begin{ruledtabular}
        \begin{tabular*}{\textwidth}{c c c c c} 
            \multicolumn{1}{c}{$Q_1$ \begin{minipage}[c][1.9cm]{\textwidth/10}\includegraphics[height=1.8cm]{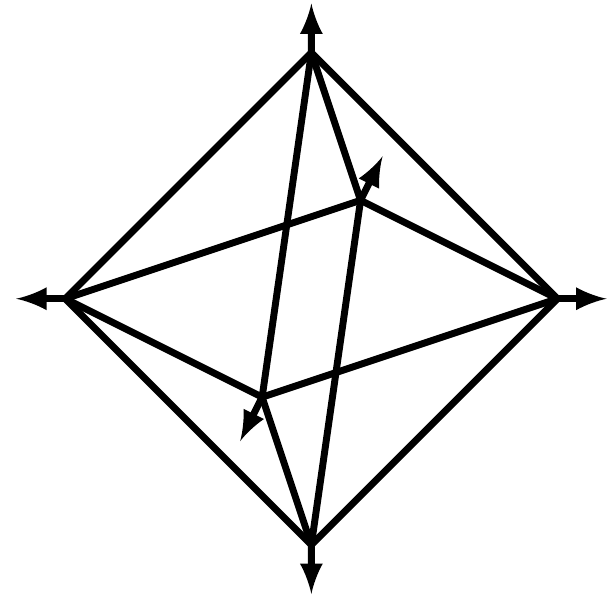}\end{minipage}} & $Q_1^\mathbf{\Gamma}$ &  $Q_1^\mathbf{R}$ & \multicolumn{1}{c}{$Q_{1\alpha}^\mathbf{X} $ \begin{minipage}[c][1.9cm]{\textwidth/10}\includegraphics[height=1.8cm]{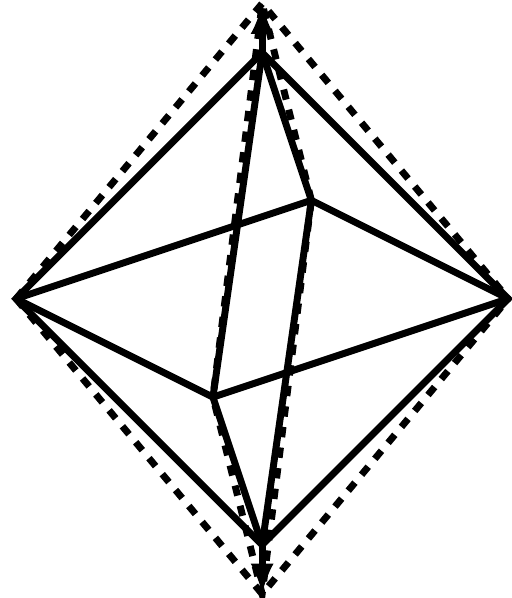}\end{minipage}} &  \multicolumn{1}{c}{$Q_{1\alpha}^\mathbf{M}$ \begin{minipage}[c][1.9cm]{\textwidth/10}\includegraphics[height=1.8cm]{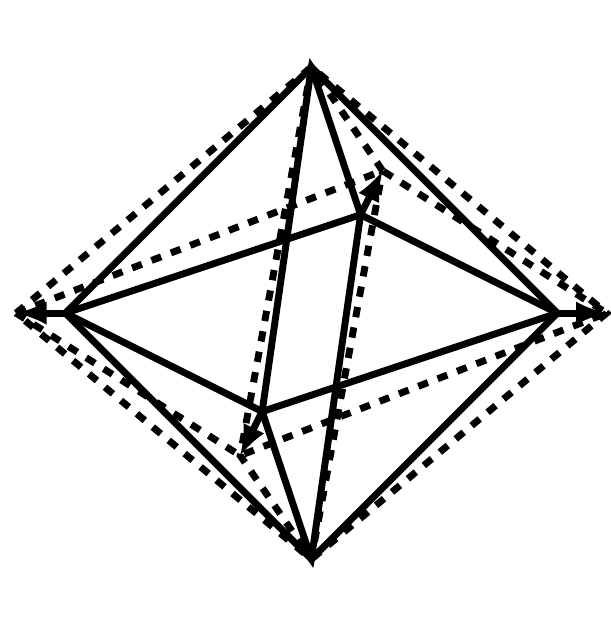}\end{minipage}} \\
            \hline
            \multirow{2}{*}{\begin{tabular}{c r}Origin in &  \textit{A}  \\ Ref. $Pm\overline{3}m$ & \textit{B} \end{tabular}} & $\Gamma_1^{+}~(a)$	&  $R_2^{-}~(a)$ & $X_3^{-}~(a,0,0)$  & $M_4^{+}~(a,0,0)$  \\
            & $\Gamma_1^{+}~(a)$ & $R_1^{+}~(a)$ &$X_1^{+}~(a,0,0)$ & $M_1^{+}~(a,0,0)$  \\ 
            \begin{minipage}[t!]{\textwidth/7}
                Displacement \\ Pattern
            \end{minipage}& \begin{minipage}[b!]{\textwidth/8}\includegraphics[width = \textwidth]{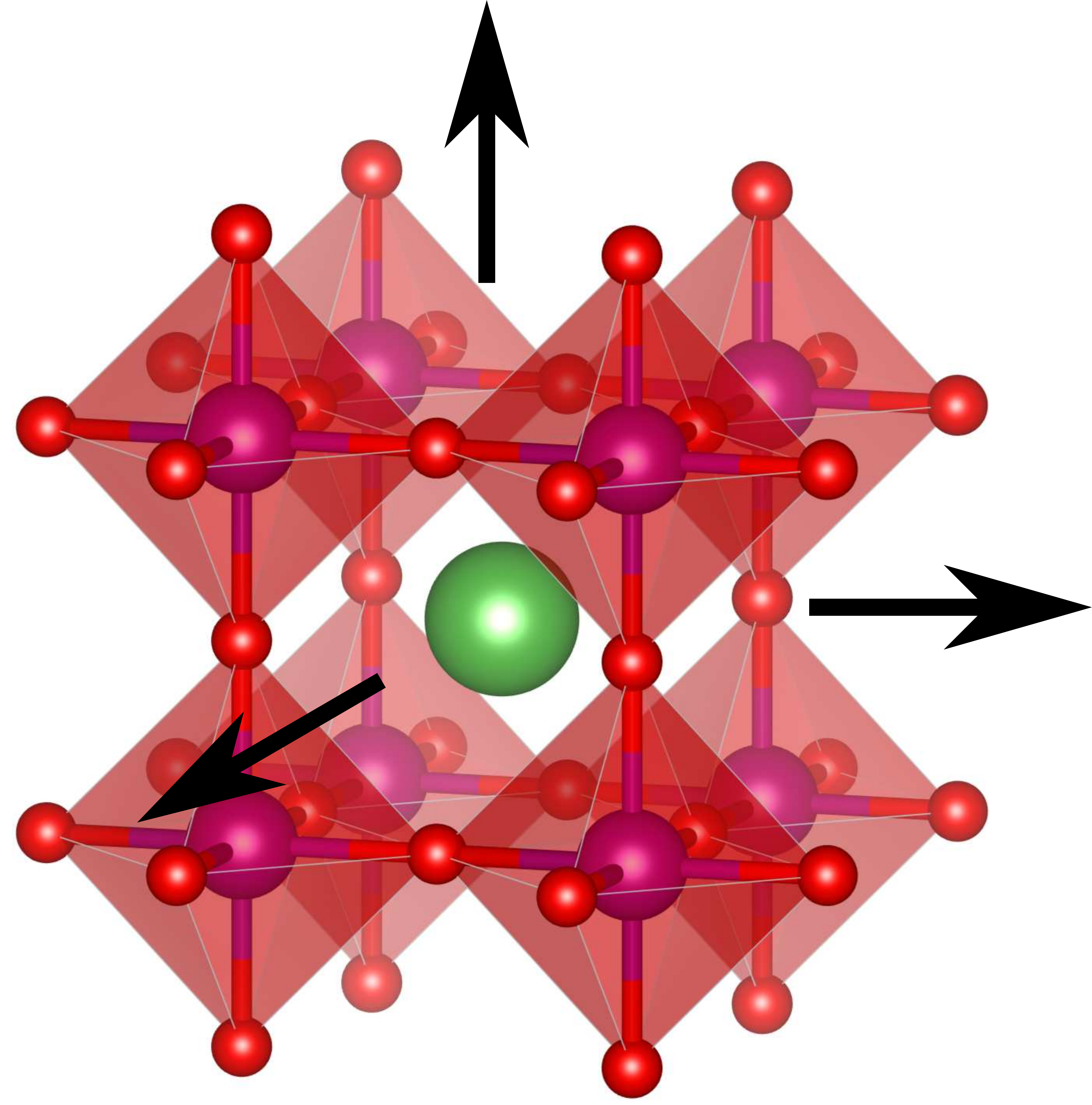}\end{minipage} & \begin{minipage}[b!]{\textwidth/8}\includegraphics[width = \textwidth]{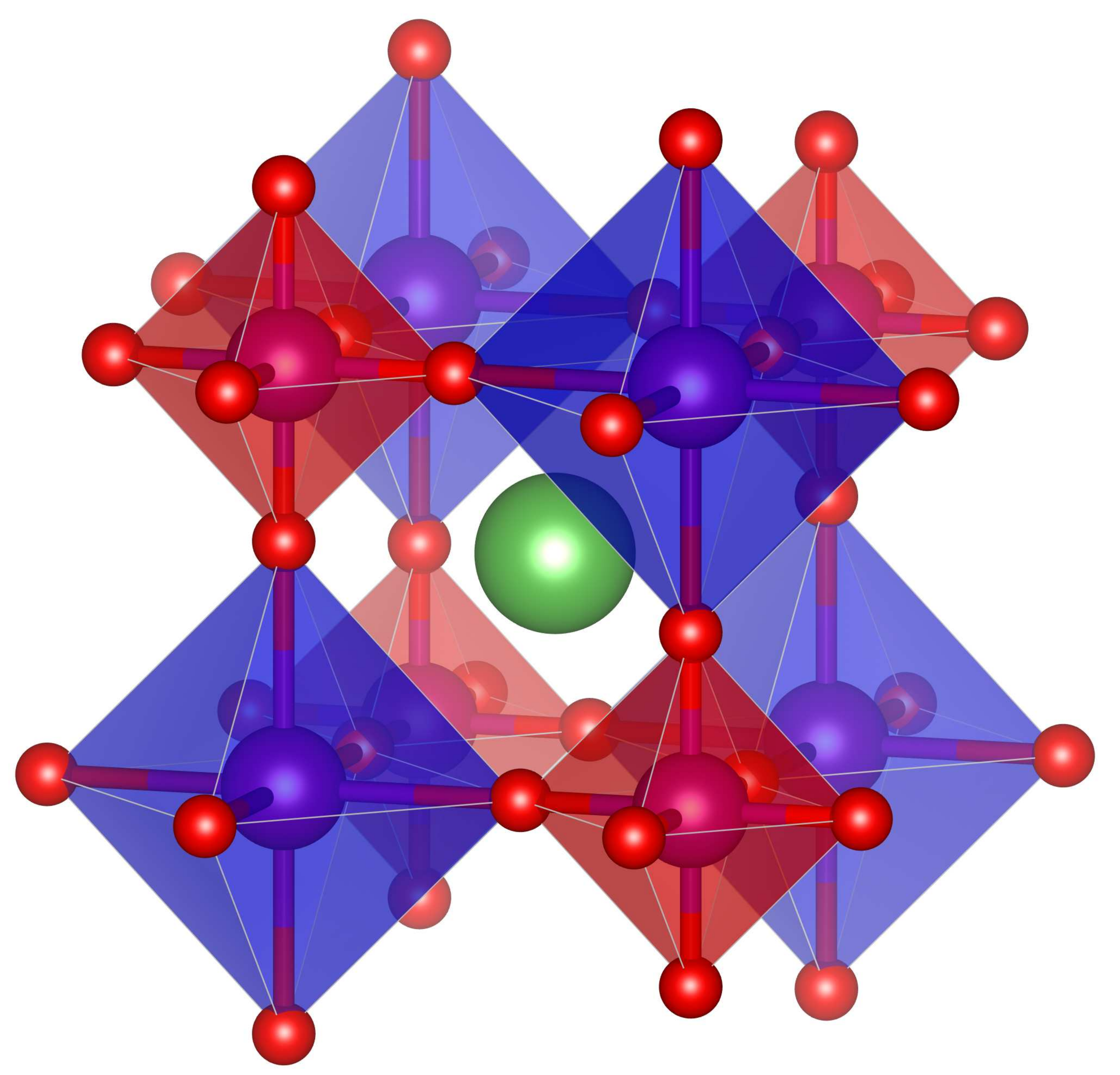}\end{minipage}& \begin{minipage}[b!]{\textwidth/8}\includegraphics[width = \textwidth]{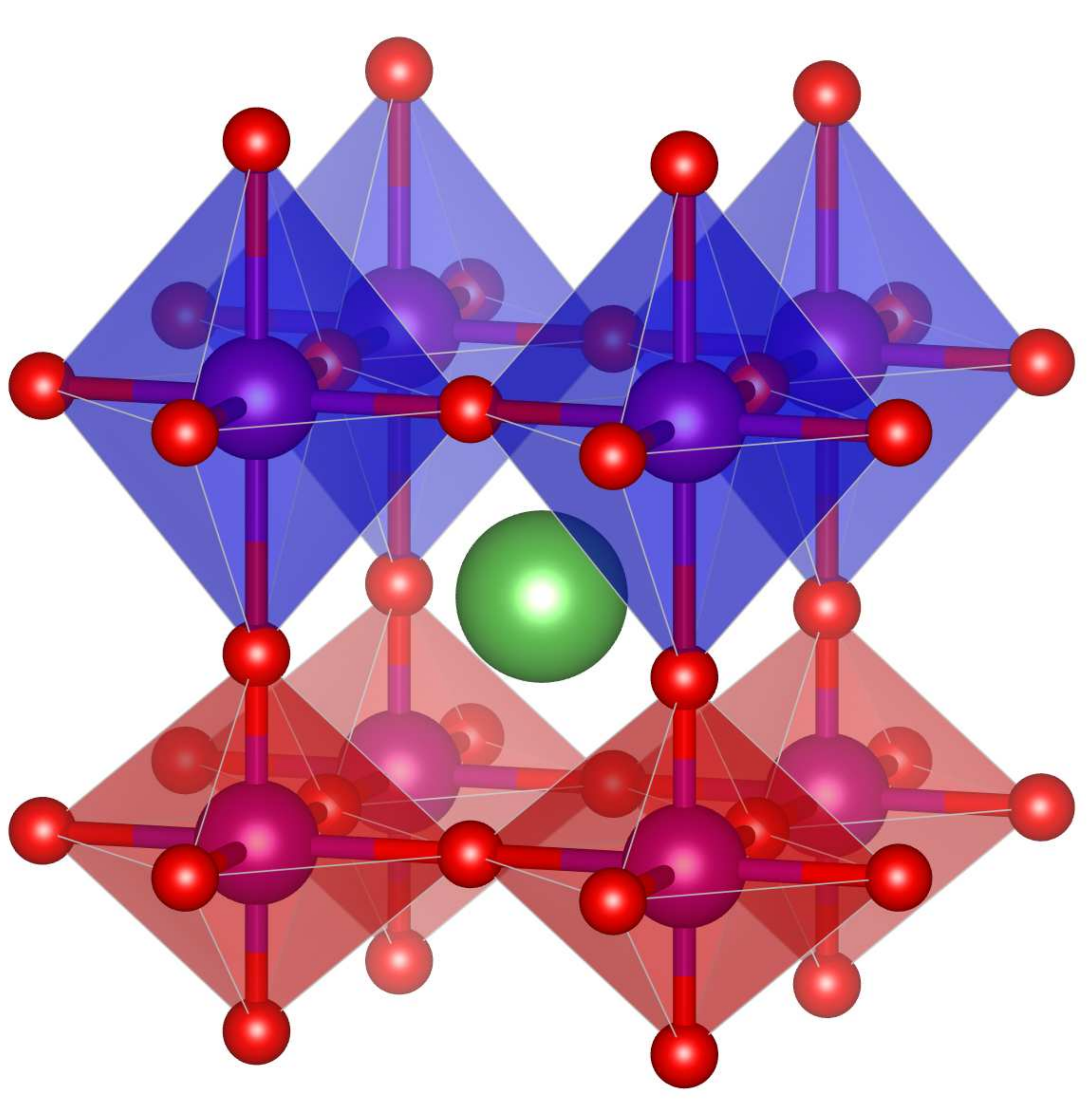}\end{minipage} & \begin{minipage}[b!]{\textwidth/8}\includegraphics[width = \textwidth]{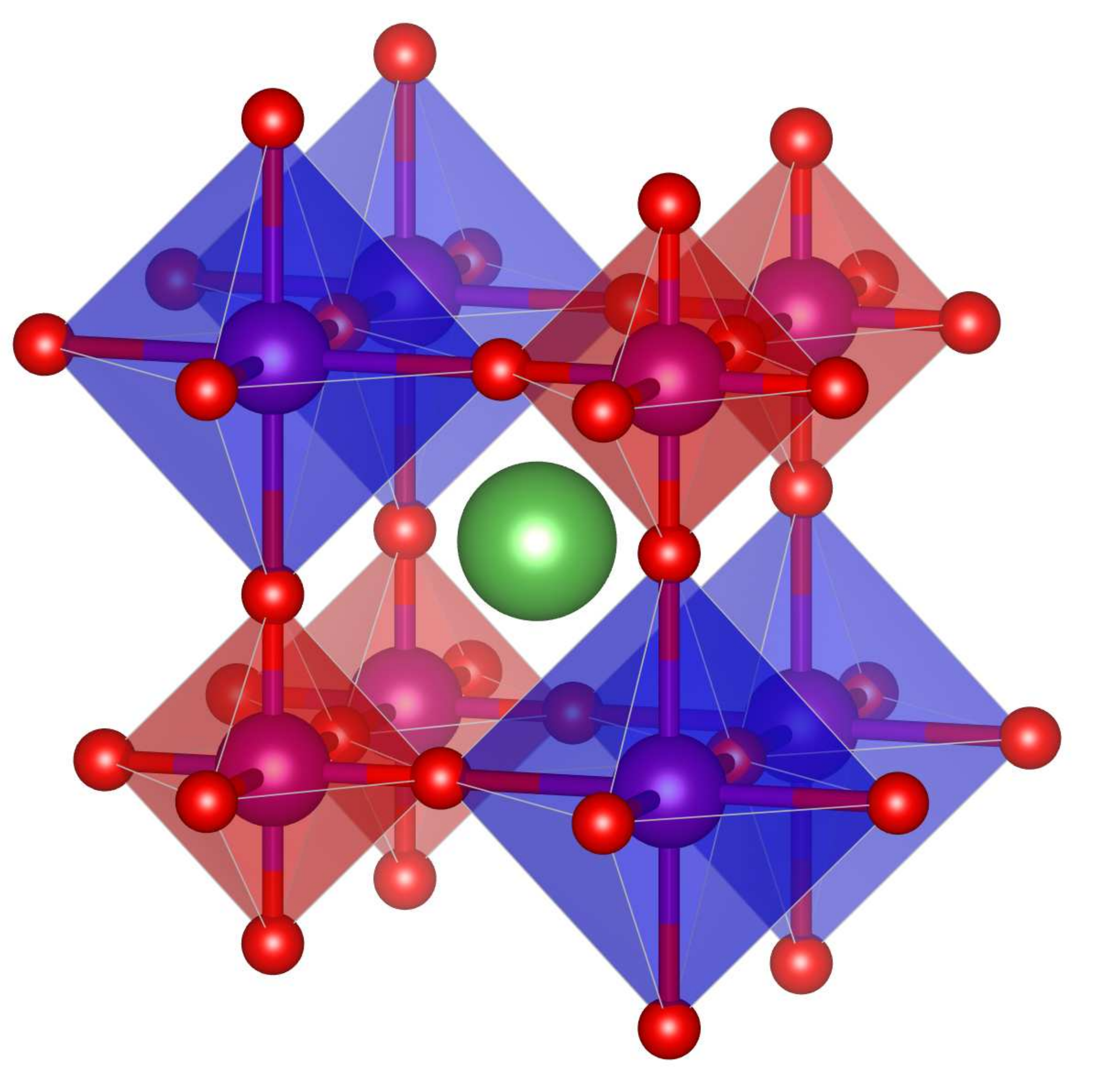}\end{minipage}  \\
            Strain Vector \vspace{0.5em}& $(a,a,a,0,0,0)$  & -  & - & - \\ 
            \begin{minipage}[t!]{\textwidth/6}
                Crystal Space Group\\ 
                (Sch\"onflies)
            \end{minipage} \vspace{0.5em} & \begin{minipage}[t!]{\textwidth/8}$Pm\overline{3}m$ \\ $(O_h^1)$ \end{minipage} & \begin{minipage}[t!]{\textwidth/8} $Fm\overline{3}m$ \\$(O_{h}^{5})$ \end{minipage} & \begin{minipage}[t!]{\textwidth/8}$P4/mmm$ \\ $(D_{4h}^{1})$ \end{minipage}  &  \begin{minipage}[t!]{\textwidth/8} $P4/mmm$ \\ $(D_{4h}^1)$ \end{minipage}  \\ 
            \begin{minipage}[t!]{\textwidth/6}
                Local Octahedral\\ 
                Symmetry
            \end{minipage} & $O_h$ & $O_{h}$ & $D_{4h}$ & $D_{4h}$ 
            \label{tab:1_Define_Q1}
        \end{tabular*}
        \begin{tabular*}{\textwidth}{c c c c c c c} 
            \multicolumn{1}{c}{$Q_2$  \begin{minipage}[c][1.9cm]{\textwidth/10} \includegraphics[height=1.8cm]{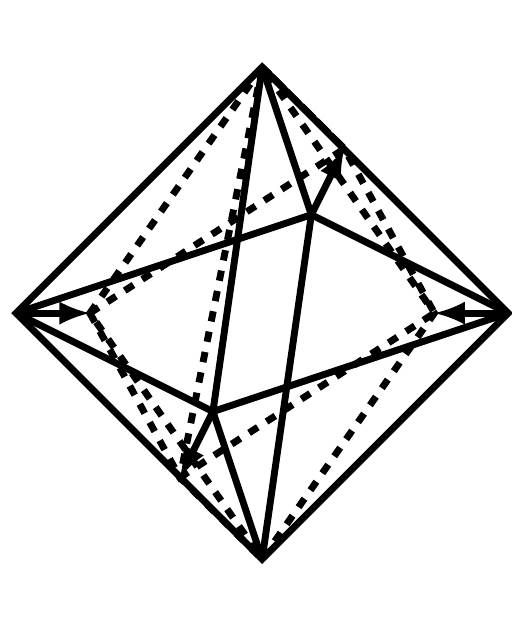}  \end{minipage}} & $Q_{2\alpha}^{\mathbf{\Gamma}} $ & $Q_{2\alpha}^{\mathbf{M}}$ & $Q_{2\alpha}^{\mathbf{R}}$& $Q_3$ \begin{minipage}[c][1.9cm]{\textwidth/11} \includegraphics[height=1.8cm]{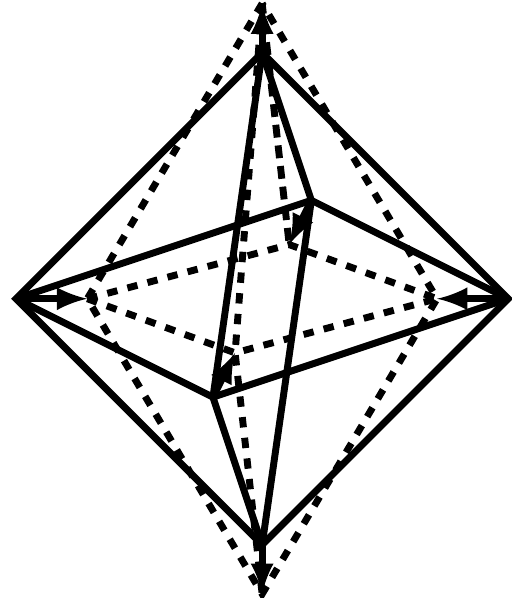}  \end{minipage} & $Q_{3\alpha}^{\mathbf{\Gamma}}$ & $Q_{3\alpha}^{\mathbf{R}}$\\
            \hline
            \multirow{2}{*}{\begin{tabular}{c r}Origin in &  \textit{A}  \\ Ref. $Pm\overline{3}m$ & \textit{B} \end{tabular}}	& $\Gamma_3^{+}~(0,a)$  & $M_3^{+}~(a,0,0)$ & $R_3^{-}~(0,a)$ & & $\Gamma_3^{+}~(a,0)$ & $R_3^{-}~(a,0)$ \\
            & $\Gamma_3^{+}~(0,a)$ & $M_2^{+}~(a,0,0)$& $R_3^{+}~(0,a)$ & & $\Gamma_3^{+}~(a,0)$ & $R_3^{+}~(a,0)$ \\ 
            \begin{minipage}[t!]{\textwidth/7}
                Displacement \\ Pattern
            \end{minipage} & \begin{minipage}[b!]{\textwidth/8}\includegraphics[width = \textwidth]{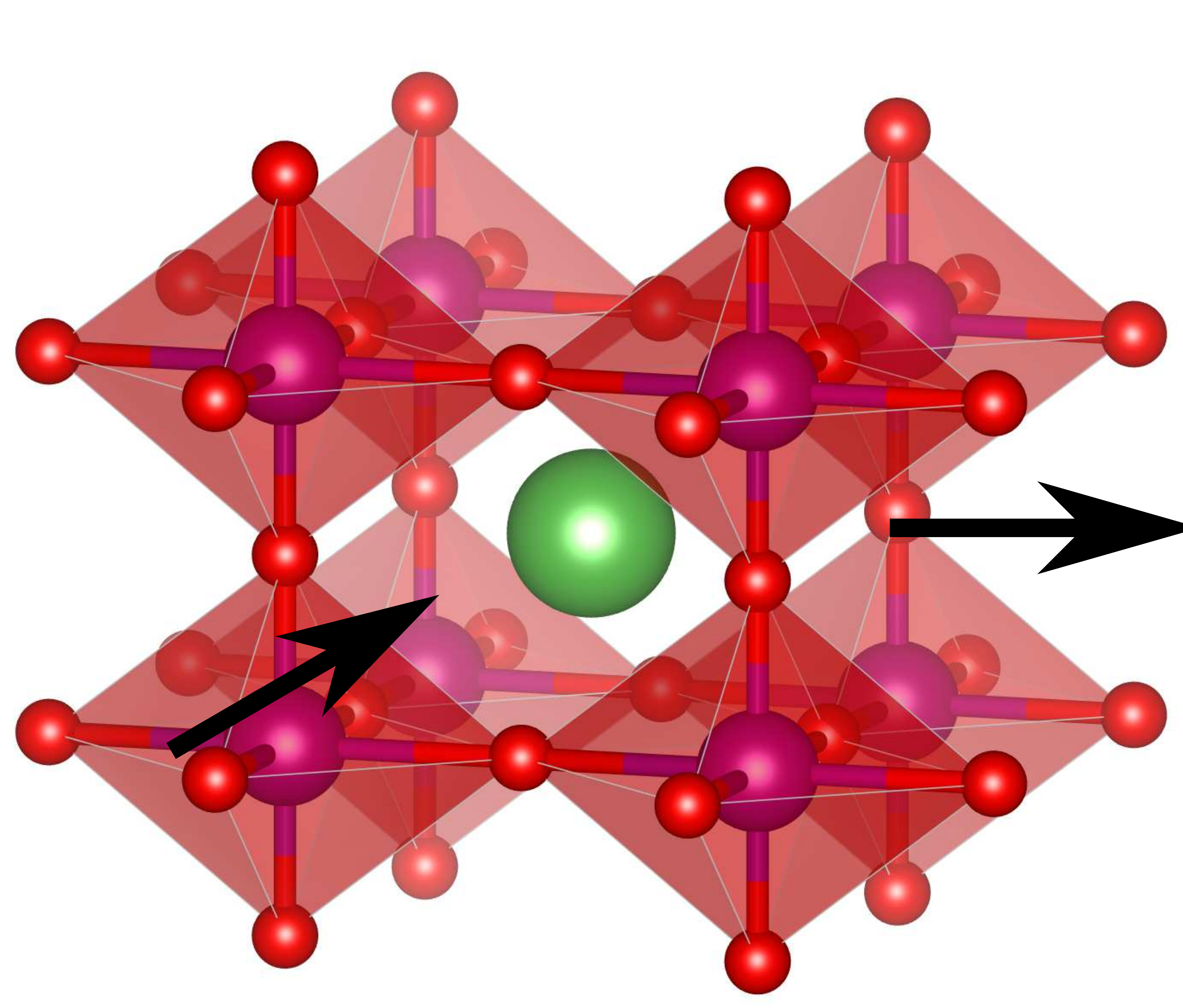}\end{minipage}& \begin{minipage}[b!]{\textwidth/8}\includegraphics[width = \textwidth]{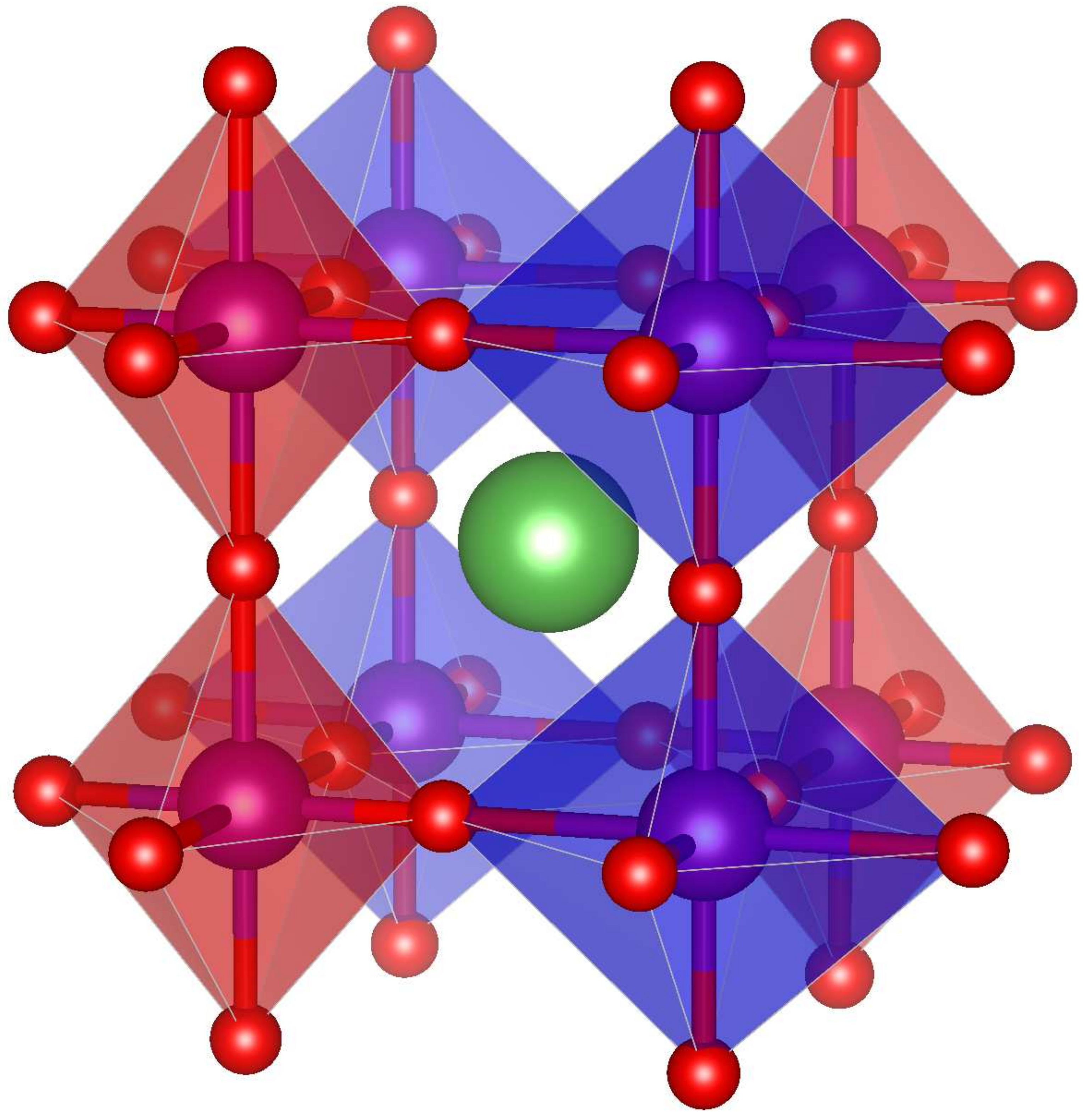}\end{minipage} & \begin{minipage}[b!]{\textwidth/8}\includegraphics[width = \textwidth]{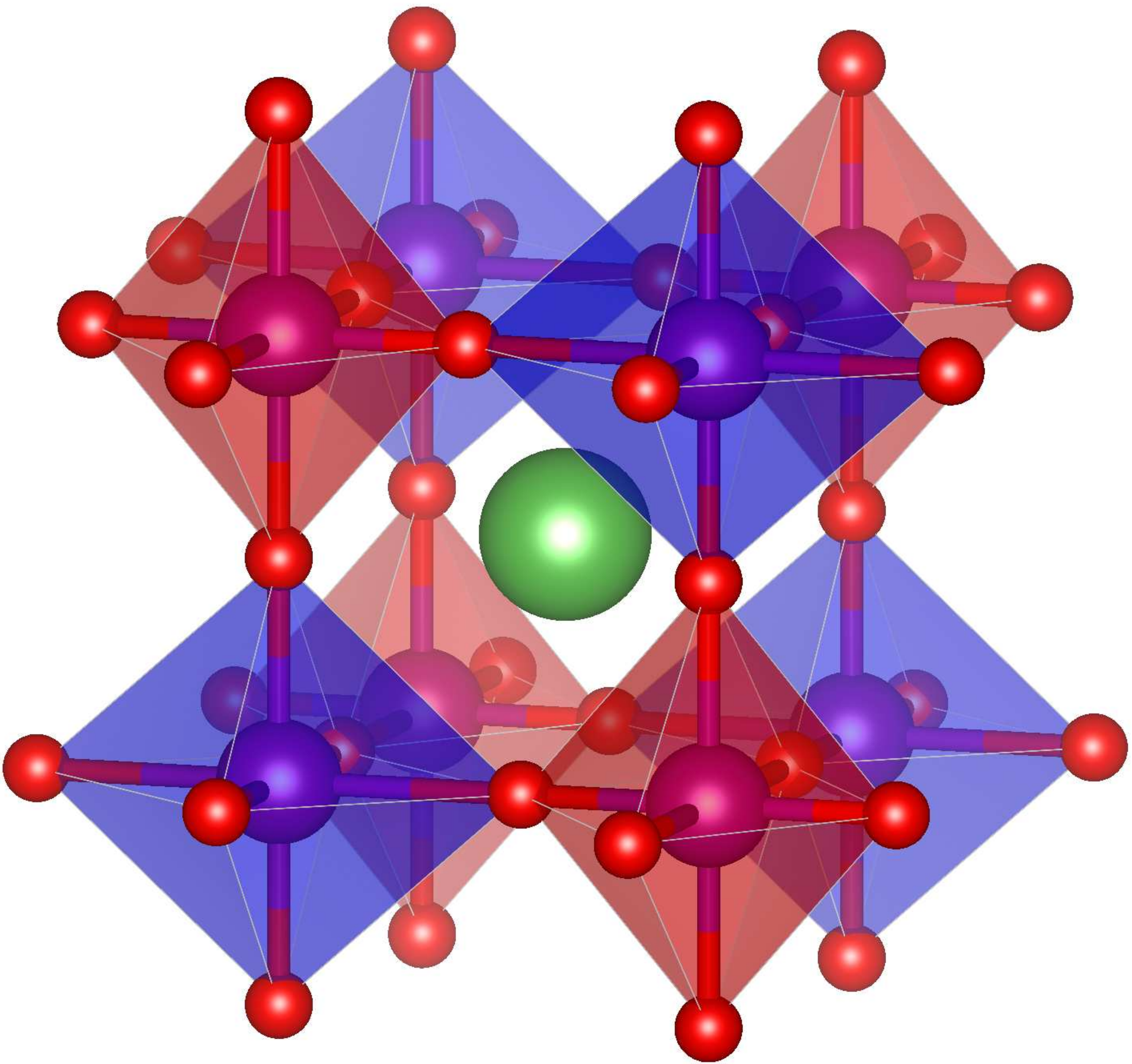}\end{minipage} & & \begin{minipage}[b!]{\textwidth/8}\includegraphics[width = \textwidth]{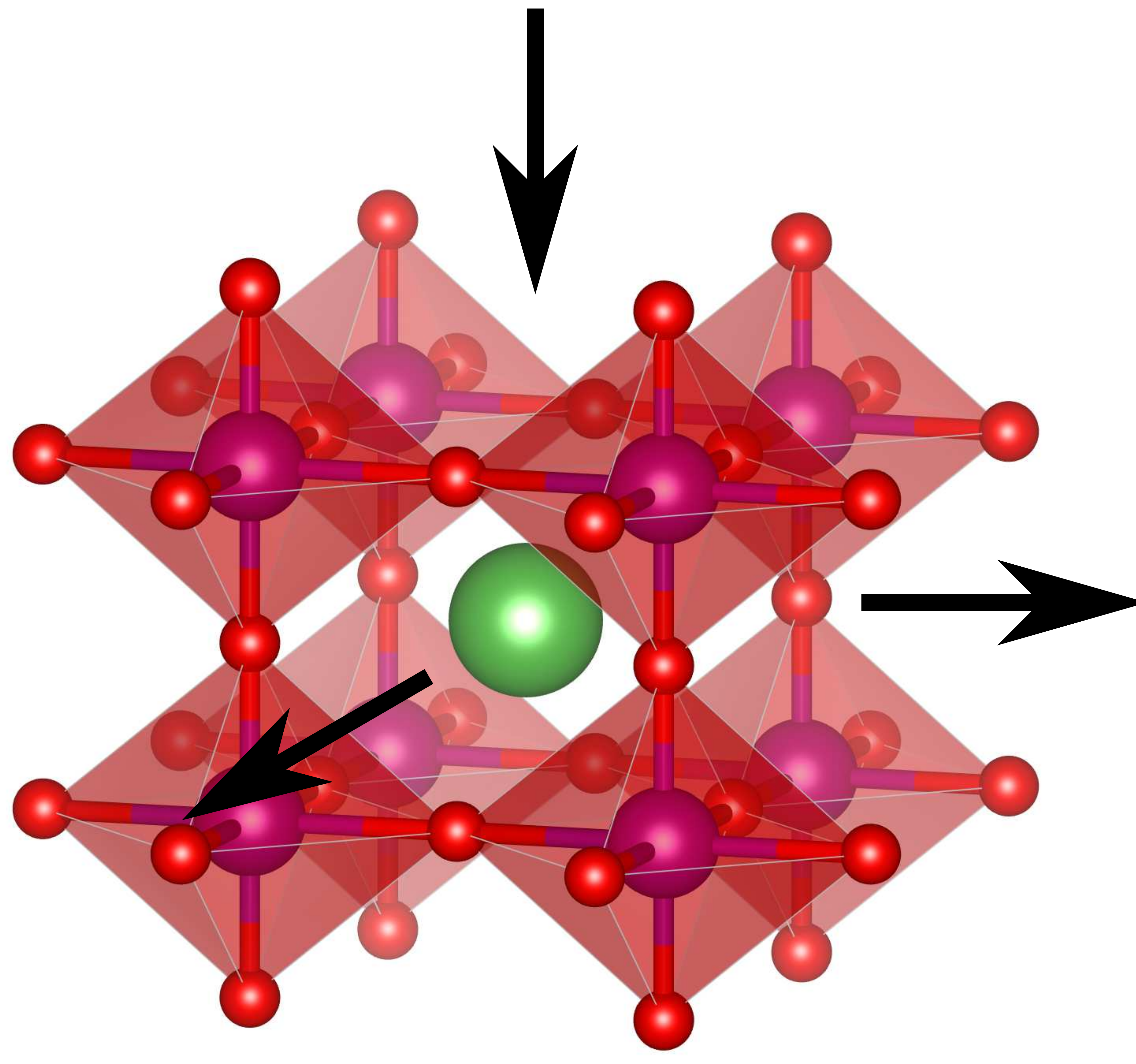}\end{minipage} & \begin{minipage}[b!]{\textwidth/8}\includegraphics[width = \textwidth]{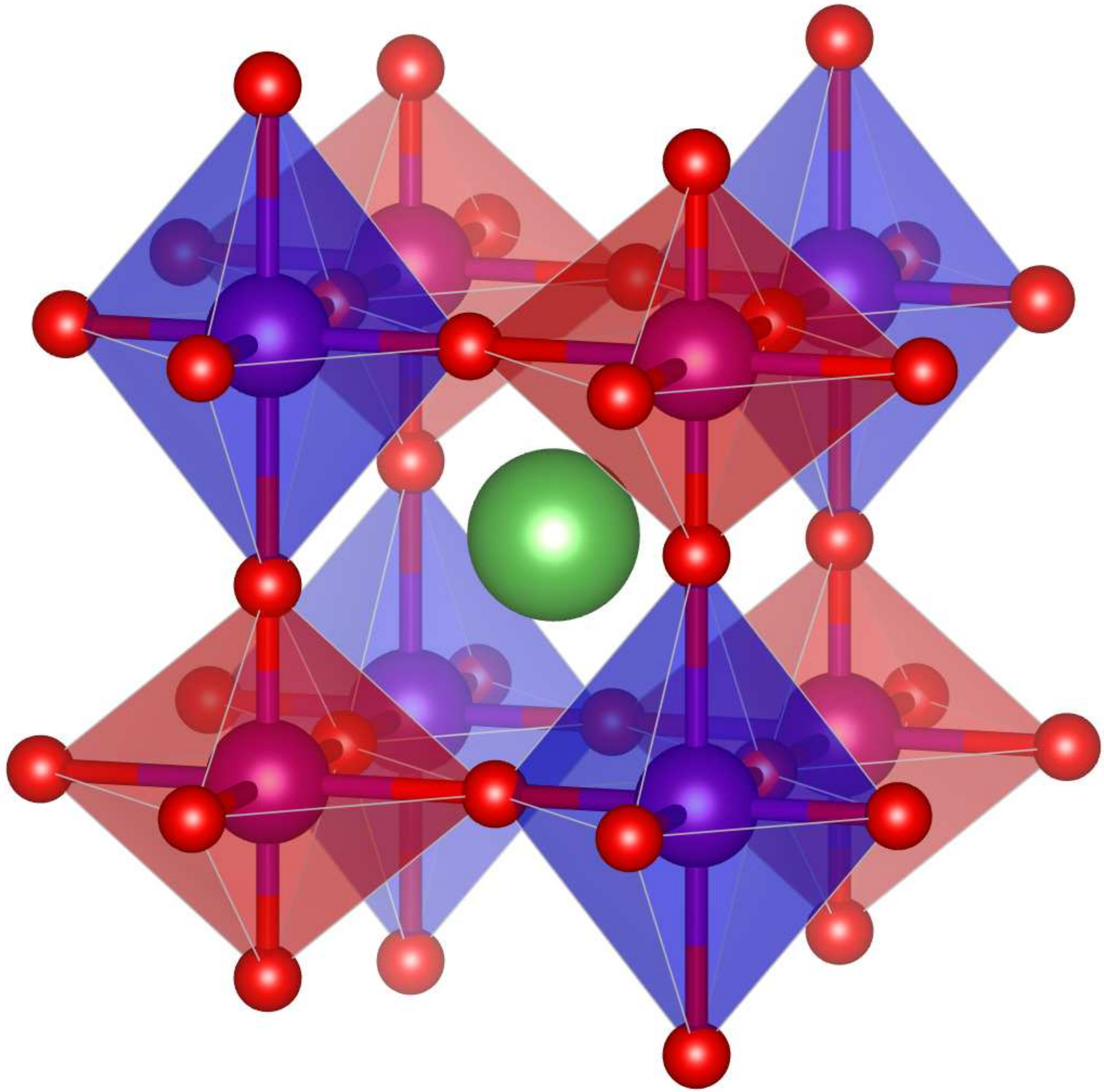}\end{minipage} \\
            Strain Vector \vspace{0.5em}& $(0,-a,a,0,0,0)$ & - & - & & $(-2a,a,a,0,0,0)$ & - \\ 
            \begin{minipage}[t!]{\textwidth/6}
                Crystal Space Group\\ 
                (Sch\"onflies)
            \end{minipage} \vspace{0.5em} & \begin{minipage}[t!]{\textwidth/8}$Pmmm$ \\ $(D_{2h}^1)$ \end{minipage}  &  \begin{minipage}[t!]{\textwidth/8} $P4/mbm$ \\ $(D_{4h}^5)$ \end{minipage} & \begin{minipage}[t!]{\textwidth/8} $I4/mcm$ \\ $(D_{4h}^{18})$ \end{minipage} & & \begin{minipage}[t!]{\textwidth/8} $P4/mmm$ \\$(D_{4h}^1)$ \end{minipage} & \begin{minipage}[t!]{\textwidth/8} $I4/mmm$ \\$(D_{4h}^{17})$ \end{minipage} \\ 
            \begin{minipage}[t!]{\textwidth/6}
                Local Octahedral\\ 
                Symmetry
            \end{minipage} & $D_{2h}$ & $D_{2h}$ & $D_{2h}$& & $D_{4h}$ & $D_{4h}$ 
        \end{tabular*}
        \begin{tabular*}{\textwidth}{c c c c  } 
            \multicolumn{1}{c}{$Q_{4,5,6}$ \begin{minipage}[c][1.9cm]{\textwidth/11}\includegraphics[height=1.8cm]{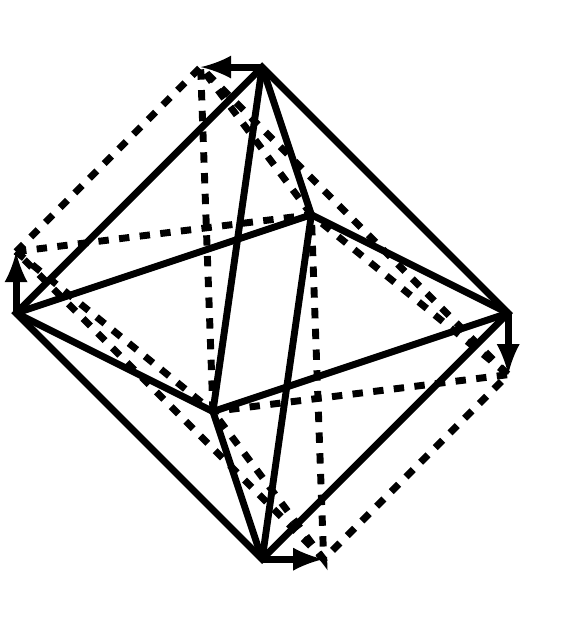}\end{minipage} } & $Q_{4\alpha}^\mathbf{\Gamma} $ & $Q_{4\alpha}^\mathbf{M}$ & $Q_{4\alpha}^\mathbf{R}$ \\
            \hline
            \multirow{2}{*}{\begin{tabular}{c r}Origin in &  \textit{A}  \\ Ref. $Pm\overline{3}m$ & \textit{B} \end{tabular}} &  $\Gamma_5^{+}~(a,0,0)$  & $M_1^{+}~(a,0,0)$ & $R_4^{-}~(a,0,0)$  \\
            & $\Gamma_5^{+}~(a,0,0)$ & $M_4^{+}~(a,0,0)$& $R_5^{+}~(a,0,0)$  \\ 
            \begin{minipage}[t!]{\textwidth/7}
                Displacement \\ Pattern
            \end{minipage} & \begin{minipage}[b!]{\textwidth/8}\includegraphics[width = \textwidth]{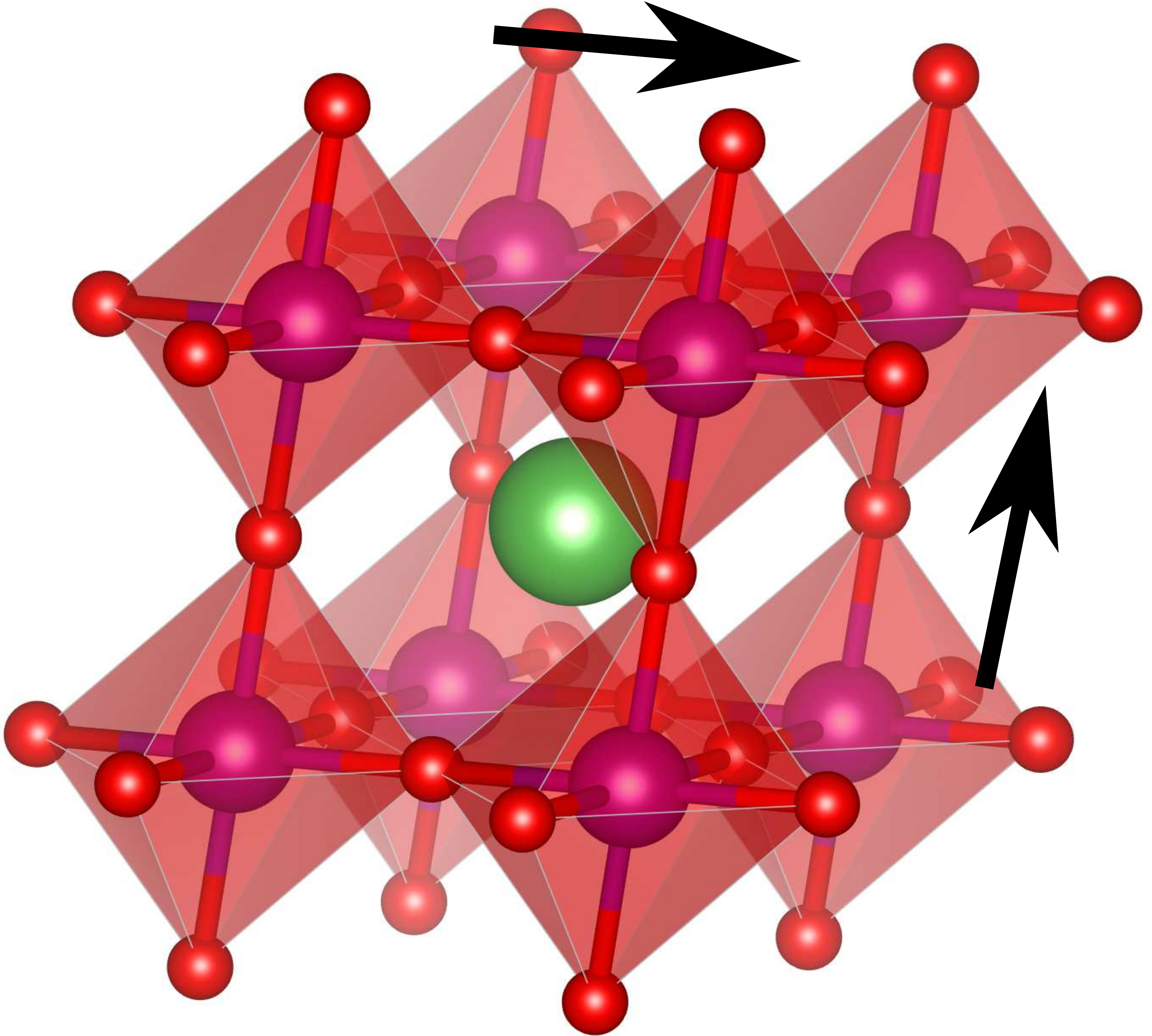}\end{minipage}& \begin{minipage}[b!]{\textwidth/8}\includegraphics[width = \textwidth]{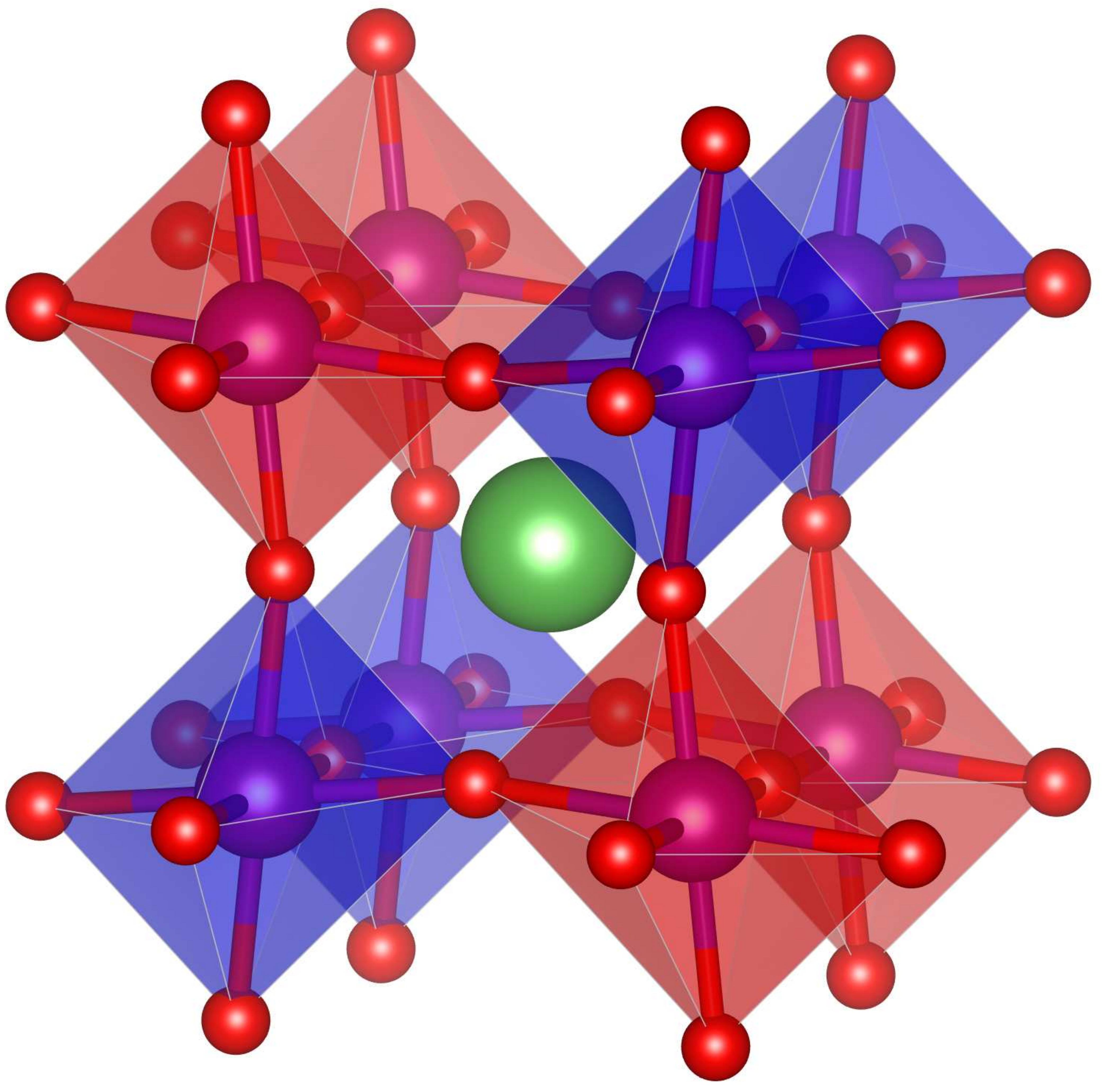}\end{minipage} & \begin{minipage}[b!]{\textwidth/8}\includegraphics[width = \textwidth]{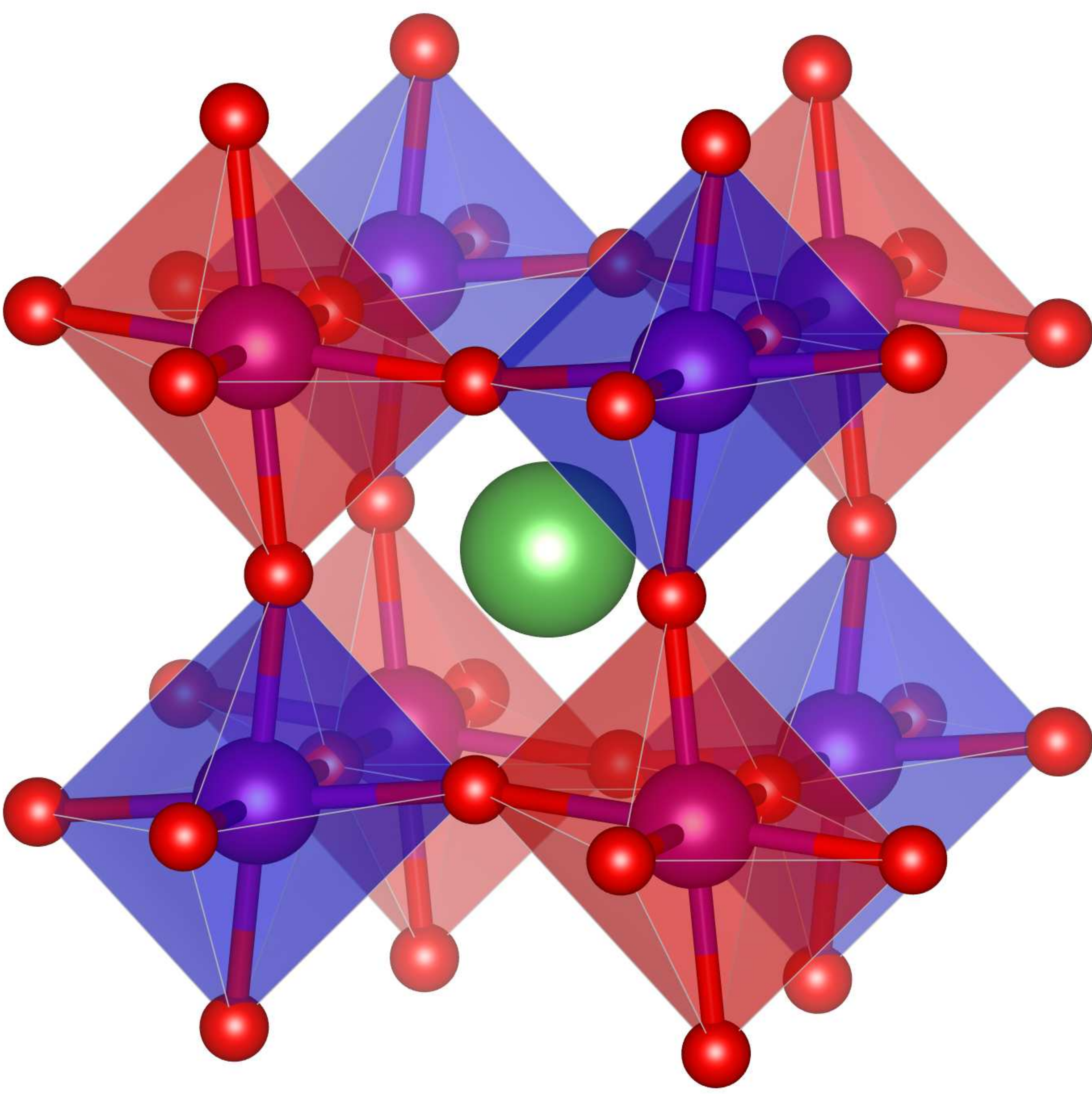}\end{minipage}  \\
            Strain Vector \vspace{0.5em} & $(0,0,0,a,0,0)$  & - & - \\ 
            \begin{minipage}[t!]{\textwidth/6}
                Crystal Space Group\\ 
                (Sch\"onflies)
            \end{minipage} \vspace{0.5em} &  \begin{minipage}[t!]{\textwidth/8}$Cmmm$ \\ $(D_{2h}^{19})$ \end{minipage}  &  \begin{minipage}[t!]{\textwidth/8} $P4/mmm$ \\ $(D_{4h}^1)$ \end{minipage} & \begin{minipage}[t!]{\textwidth/8} $I4/mmm$ \\$(D_{4h}^{17})$ \end{minipage} \\ 
            \begin{minipage}[t!]{\textwidth/6}
                Local Octahedral\\ 
                Symmetry
            \end{minipage}  & $D_{2h}$ & $D_{2h}$ & $D _{2h}$ 
        \end{tabular*}
    \end{ruledtabular}
\end{table*}

Regardless, in perovkites with interconnected Jahn-Teller centers, it is usual to quantify  the amplitude of $Q_2$ and $Q_3$ distortions based on $B{-}O$ distances in absolute coordinates. This notation quantifies the distortion of one \textit{individual} octahedron. It does not indicate the cooperative arrangement of the distorted octahedra nor distinguish condensed phonon-type distortions from homogeneous lattice strain. At the same time, the quantification and notation of $Q_4~-~Q_6$  distortions seems to have been dropped in latter years (the last occurrence we found dates back to 1997\cite{Bacci1997}).

\textit{Carpenter} and \textit{Howard} gave a different notation based on the \textsc{isotropy} software suite and associating Jahn-Teller ordering schemes with labels of irreducible representations (irrep) and ordering parameters \cite{Carpenter2009}. These symmetry labels are unique and distinguish between strain and phonon modes. Moreover, the symmetry-adapted analysis allows to quantify the amplitudes of Jahn-Teller distortions in their own subspace, such that they can be separated from other distortions in the crystal lattice as octahedral rotations or antipolar motions. Finally, by creating invariant polynomial terms between the subspace of the Jahn-Teller distortions and other lattice distortions, the order, sign and strength of couplings between those different distortions can be studied. This makes the decomposition of lattice distortions into orthogonal irreducible subspaces a very powerful approach.  However, the application of the symmetry analysis has not found widespread application. A reason might be that the connection between the \textit{Van-Vleck}-numbering and the irrep labels is not obvious.
 
In the context of a first-principles study of RNiO$_3$ rare-earth nickelates, \textit{He} and \textit{Millis} \cite{He2015} defined labels $Q_x^k$ (which could be said to be inspired by \textit{Kanamori} \cite{Kanamori1960}) with $x$ a number indicating a local pattern (different from \textit{van Vleck's}) and $k$ the label associated to high symmetry k-points in the cubic Brillouin zone. Through the phase factor $e^{i\vec{k}\vec{x}}$, the k-label emphasizes the cooperative arrangement. However, they only labeled the modes of interest in their study, without labeling all possibilities.

Here, we introduce canonical notations defining a unique symbol for all possible cooperative Jahn-Teller distortions in the perovskite structure.
Our canonical symbols have the form  $Q_{i\alpha}^{\vec{q}}$. The subscript \emph{i} indicates the local distortion pattern and takes the enumeration of the octahedral normal modes from \textit{Van-Vleck}. The second subscript $\alpha$ is necessary for non-isotropic local patterns that break the cubic symmetry of the octahedra (all besides $Q_1$): it identifies the alignment of the unique feature of the local distortion pattern with respect to the perovskite lattice. It takes the values $x,y,z$, which are defined to lie along the cubic perovskite lattice axes. For a two-dimensional local distortion pattern, the unique feature is the axis orthogonal to the two-dimensional distortion plane (applies to $Q_2$ and $Q_4$). For a one or three dimensional local distortion pattern it shows the cartesian axis along the unique feature.  
Finally, the superscript $\vec{q}$ is the label of the reciprocal space vector according to which the local mode is translating in the crystal. Within this work, we limit $\vec{q}$ to zone center ($\mathbf{\Gamma} = (0,0,0)$) and zone boundary modes at high symmetry $\vec{q}$ points. The zone center $\mathbf{\Gamma}$ is thereby associated to lattice strains. However, there is no inherent limitation of the notation to the high symmetry $\vec{q}$ - points. 
In the cubic Brillouin zone, the high-symmetry $\vec{q}$ points at the zone boundary are $\mathbf{X}=($\textonehalf$,0,0),\mathbf{M}=($\textonehalf,\textonehalf$,0)$, and $\mathbf{R} = ($\textonehalf,\textonehalf,\textonehalf$)$. The power of using such high-symmetry $\vec{q}$ points lies in their unique definition of the cooperative arrangement of the local distortion pattern and thereby also the related orbital ordering.  In analogy to magnetic orderings, $\mathbf{\Gamma}$ leads to \emph{ferro}, $\mathbf{X}$ to a planar or \emph{A-type}, $\mathbf{M}$ to a columnar or \emph{C-type}, and $\mathbf{R}$ to a checkerboard or \emph{G-type} arrangement. The freedom of the phase-factor depends on the local distortion pattern, since the corner-shared atoms imply the opposite displacement of neighboring octahedra. The resulting notations for all local patterns at the high symmetry points are shown in Table \ref{Tab:Define_Qi}. Additionally, Table \ref{Tab:Define_Qi} shows the crystal symmetry achieved by condensing the individual cooperative modes in the $Pm\overline{3}m$ space-group, the local octahedral symmetry only taking into account the MX$_6$ complex, and the label of the irreducible subspace with the origin of the cubic perovskite unit cell set on either \textit{A} or \textit{B} cation.

The $Q_1$ mode is related to a homogeneous expansion/contraction of the volume of individual octahedra. It appears as a lattice strain at $\mathbf{\Gamma}$. As in the molecular case, it can be omitted by choosing a reference stationary with respect to $Q_1^\mathbf{\Gamma}$. Since the local distortion pattern is three dimensional, $Q_1$ is limited to $\vec{q}$ between $\mathbf{\Gamma}$ and $\mathbf{R}$. $Q_1^{\mathbf{R}}$ is often called the breathing type distortion and associated to charge ordering \cite{Mercy2017,Balachandran2013}. Two additional modes changing the volume of local octahedra can be thought of. First, a mode that alters one bond axis (uniaxial volume change ) and, second, two octahedral axis (planar volume change). In the molecular case, these distortions do not appear as normal modes as they are not orthogonal to $Q_1$ and $Q_3$. Since in solids these modes have been shown to be connected to charge ordering \cite{Park2017}, we associate equally a $Q_1$-label to them. In the periodic perovskite crystal, the uniaxial volume change appears as a irreducible mode at $\mathbf{X}$ ($Q_{1\alpha}^{\mathbf{X}}$) and the planar volume change at $\mathbf{M}$  ($Q_{1\alpha}^{\mathbf{M}}$ in Table \ref{Tab:Define_Qi}). At the other high-symmetry q-points of the cubic Brillouin zone, the uniaxial and planar volume changes are (equivalently to the case of an isolated octahedron) not orthogonal to the other modes presented in Table \ref{Tab:Define_Qi}. Indeed, at the $\mathbf{M}$ point, the uniaxial volume change is represented by a sum of $Q_{1\alpha}^\mathbf{M}$ and $Q_{2\alpha}^\mathbf{M}$ while, at  the $\mathbf{R}$  and $\mathbf{\Gamma}$ points, the volume changes are represented  by sums of $Q_{1}^\mathbf{R/\Gamma}$, $Q_{2\alpha}^\mathbf{R/\Gamma}$ and $Q_{3\alpha}^\mathbf{R/\Gamma}$ (respectively the subspaces $R_2^-/R_3^-$ and $\Gamma_1^+/\Gamma_3^+$ ).

The $Q_2$ modes are two-dimensional and can hence translate with $\mathbf{\Gamma}$,$\mathbf{M}$, and $\mathbf{R}$. They reduce the local symmetry to $D_{2h}$ stabilizing a mixed $d_{z^2-r^2}/d_{x^2-y^2}$ state. 

The $Q_3$ modes are three-dimensional and hence appear at $\mathbf{\Gamma}$ and $\mathbf{R}$. They reduce the local symmetry to $D_{4h}$ stabilizing, either a $d_{x^2-y^2}$ or a $d_{z^2-r^2}$ state for tetragonal compression or extension respectively. 

At $\mathbf{\Gamma}$ and $\mathbf{R}$, $Q_2$ and $Q_3$ form a two-dimensional subspace equivalent to the $Q_2/Q_3$ space of the isolated Jahn-Teller center. However, an intriguing difference to the isolated center is the appearance of $Q_{2\alpha}^\mathbf{M}$ in its own subspace. This gives an additional degree of freedom for cooperative Jahn-Teller distortions of connected Jahn-Teller centers.
 
Finally, the $Q_4$ modes label the shear distortions. As they are two-dimensional, they appear at $\mathbf{\Gamma}$,$\mathbf{M}$, and $\mathbf{R}$, being at each point threefold degenerated. This threefold degeneracy reflects the modes $Q_5$ and $Q_6$ in \emph{Van Vleck's} numbering. The necessity of $Q_5$ and $Q_6$ falls away using the second subscript $\alpha$ in our notations. $Q_4$ modes reduce the local symmetry to $D_{2h}$ albeit in a different way than $Q_2$ since the $B-O$ distances in the sheared plane stay degenerate.

\begin{figure}
    \includegraphics[width=\columnwidth]{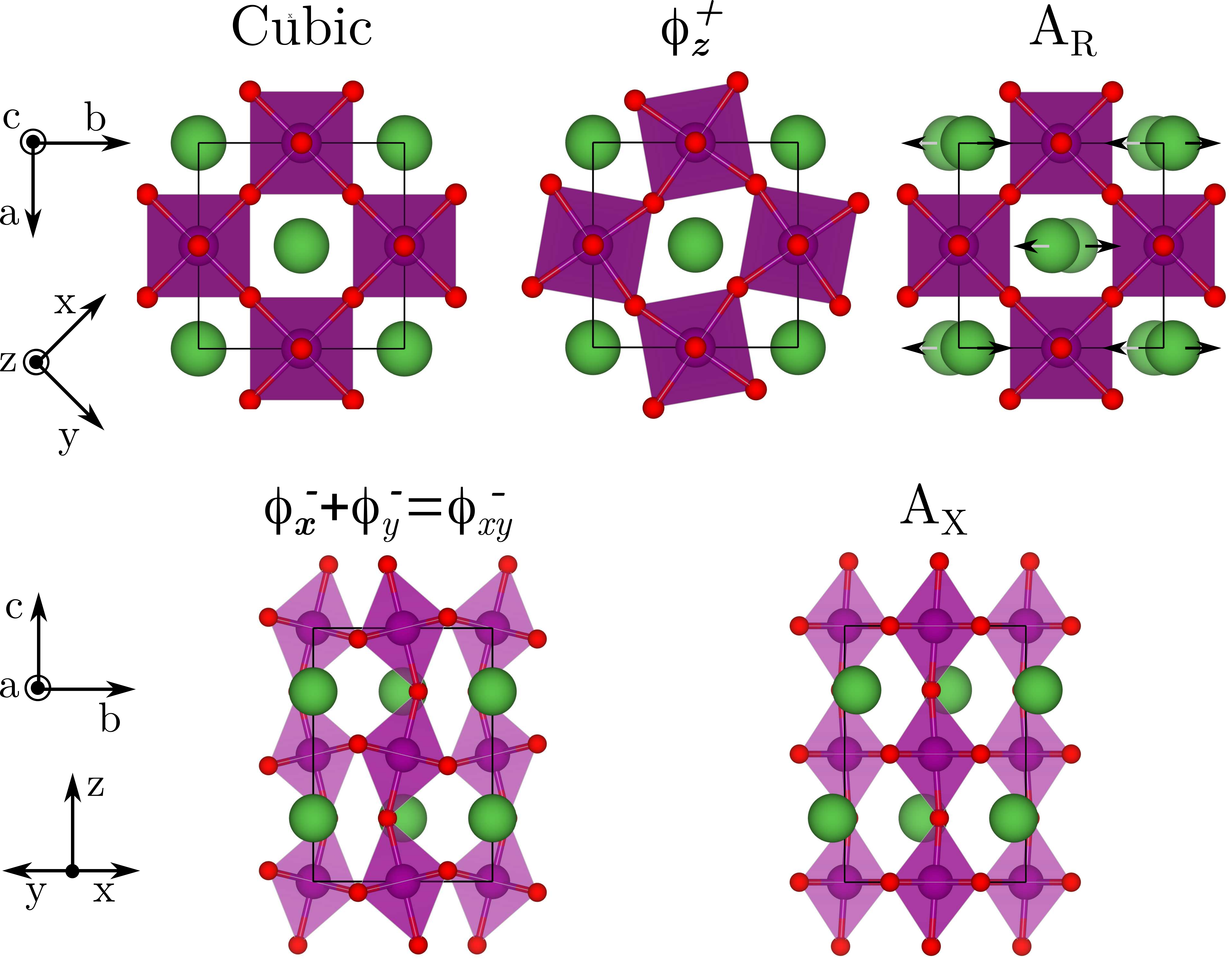}
    \caption{Displacement patterns of condensed symmetry adapted modes in the LaMnO$_3$ \textit{Pbnm}-phase (excluding Jahn-Teller distortions). Cubic \textit{xyz}- and orthorhombic \textit{abc}-coordinate system used throughout the paper are indicated. The \textit{Pbnm}-unit cell is shown by the black continuous line. a) Reference cubic positions, b) in-phase rotation $\phi_z^+$ (irrep: $M_3^+$), c) antipolar motion $A_R$ at the R-point of the cubic Brillouin zone (irrep: $R_4^-$), d) out-of-phase rotations $\phi_{xy}^-$ (irrep: $R_5^-$), e) antipolar motion $A_X$ at the X-point of the cubic Brillouin zone (irrep: $X_5^-$).}
    \label{fig:modes} 
\end{figure}

All irreducible subspaces besides $X_3^-/X_1^+$ and $R_4^-/R_5^+$ given in Table \ref{Tab:Define_Qi} are formed exclusively by the corresponding Jahn-Teller displacements of the ions at the octahedral corners. In the subspaces  $X_3^-/X_1^+$ and $R_4^-/R_5^+$ additional antipolar motions of \textit{A}-cations are found. In $X_3^-/X_1^+$ the $A$-cations of $[100]$-planes move along the corresponding cubic axes. In the $R_4^-/R_5^+$ subspace, it is the case for  $A$-cations of the $[111]$-planes (see also Fig. \ref{fig:modes}).  Hence, it is expected that the condensation of a $Q_1^\mathbf{X}$ or $Q_4^\mathbf{R}$ distortion will induce the corresponding antipolar motion and vice versa.

Finally we notice that the strains $Q_1^\mathbf{\Gamma}$,$Q_{2\alpha}^\mathbf{\Gamma}$,$Q_{3\alpha}^\mathbf{\Gamma}$, and $Q_{4\alpha}^\mathbf{\Gamma}$ represent a complete strain basis for the cubic perovskite system.

These canonical notation, defining a unique symbol for each cooperative Jahn-Teller distortion while distinguishing phonon-modes and lattice strains, will facilitate the discussion of perovskite systems experiencing static Jahn-Teller distortions. As it will be shown in the forthcoming of this work, the orthogonality of the decomposition is most powerful in the study of the interplay of Jahn-Teller distortions with other lattice distortions and strains.  

\section{Ground State Properties}
\label{sec:IV_GS} 

In this section we review the structural, magnetic, and dielectric properties of the LaMnO$_3$ bulk ground-state phase. We compare the results of our DFT+$(U|J)$ calculations to experimental values in order to assess the validity of our calculation method (See Table \ref{tab:2_all_prop}). 
 
The ground-state \textit{Pbnm} phase can be described in terms of its atomic distortion with respect to the aristotype cubic perovskite structure, taken as reference. This distortion can be decomposed according to the orthorgonal symmetry-adapted phonon modes and lattice strains defined by the irreducible representations of the cubic reference structure. The modes with the largest amplitudes are (i) one in-phase rotation of the oxygen octahedra ($\phi_z^+$ irrep: $M_2^+$) and (ii) two out-of-phase rotations ($\phi_x^- + \phi_y^- = \phi_{xy}^-$ irrep: $R_5^-$), leading together to the $a^-a^-c^+$ rotation pattern \cite{Glazer1972} and reducing the symmetry to the \textit{Pbnm} space group. This rotation pattern further induces two antipolar motions of the La cations\cite{Varignon2015,Miao2016,Miao-2013}. Firstly, an antipolar motion of the La atoms (and also oxygens) of consecutive $(001)$-planes along the pseudocubic $xy$-direction ($A_X$ irrep: $X_5^-$). Secondly, an antipolar motion  of La atoms of consecutive (111)-planes equally along the pseudocubic $xy$-direction ($A_R$ irrep: $R_4^-$). This latter antipolar motion possesses the same irrep as the Jahn-Teller modes $Q_{4\alpha}^R$ defined in Table \ref{Tab:Define_Qi}. The respective oxygen motions $Q_{4x}^{\mathbf{R}}$ and $Q_{4y}^{\mathbf{R}}$ do appear also but with an amplitude one order of magnitude smaller than the already small amplitude of the $A_R$ cation motions so that they are not reported in Table \ref{tab:2_all_prop}. Finally, the ground-state structure also shows a significant Jahn-Teller distortion $Q_{2z}^{\mathbf{M}}$, and sizable tetragonal (compressive along z-axis) and shear strains $Q_{3z}^{\mathbf{\Gamma}}$ and $Q_{4z}^{\mathbf{\Gamma}}$, all compatible with the \textit{Pbnm} symmetry. 
The atomic displacement patterns associated to these modes (excluding the strains and Jahn-Teller modes, already sketched in Table \ref{Tab:Define_Qi}) are shown in Fig. \ref{fig:modes}.


\begin{table}
    \begin{threeparttable}
        \renewcommand{\arraystretch}{1.35}
        \caption{Comparison of quantities calculated from DFT with PBEsol+($5\lvert 1.5$) and PBEsol+($8\lvert 2$) with experimental values. Top: amplitudes of the symmetry-adapted modes (\si{\angstrom}) extracted with \textsc{isodistort}\tnote{a} of relaxed LaMnO$_3$ with imposed AFM-A magnetic order. Center: optical dielectric permittivity tensor $\epsilon^{\infty}$ and electronic band gap $E_{Gap}$ (eV). Bottom: magnetic moment $\mu$ ($\mu_B$), magnetic exchange constants $J$ (meV) and Neel-Temperature $T_N$ (K).}
        \begin{ruledtabular}
            \begin{tabular*}{\columnwidth}{cccc}
                &  ($5\lvert 1.5$) &     ($8\lvert 2$)   &  Exp.  \\    
                \hline 
                \multicolumn{4}{c}{\textbf{\underline{Structure}}}\\
                
		    \multirow{2}{*}{ \begin{tabular}{c} $Q_{4z}^\mathbf{\Gamma}$ \\  $\Gamma_5^{+}~(a,0,0)$ \end{tabular}} & \multirow{2}{*}{-0.036}  & \multirow{2}{*}{-0.039}  & \multirow{2}{*}{-0.027\tnote{b}/-0.027\tnote{c}} \\ 
                & & &  \\
                
		    \multirow{2}{*}{ \begin{tabular}{c}  $Q_{3z}^\mathbf{\Gamma}$ \\ $\Gamma_3^{+}~(a,0)$ \end{tabular} } &   \multirow{2}{*}{-0.040} & \multirow{2}{*}{-0.040} & \multirow{2}{*}{-0.032\tnote{b}/-0.032\tnote{c}} \\         	 
                & & &  \\
                
		    \multirow{2}{*}{ \begin{tabular}{c} $A_X$ \\ $X_5^{-}~(0,0,0,0,a,-a)$ \end{tabular} } & \multirow{2}{*}{0.33}  & \multirow{2}{*}{0.34} &  \multirow{2}{*}{0.30\tnote{b}/0.29\tnote{c}} \\
                & & &  \\
                
		    \multirow{2}{*}{ \begin{tabular}{c} $\phi_z^+$ \\ $M_2^{+}~(a,0,0)$ \end{tabular} }  & \multirow{2}{*}{0.49}  & \multirow{2}{*}{0.51}   & \multirow{2}{*}{0.48\tnote{b}/0.48\tnote{c}} \\
                & & &  \\
                
		    \multirow{2}{*}{ \begin{tabular}{c} $Q_{2z}^{\mathbf{M}}$ \\ $M_3^{+}~(a,0,0)$ \end{tabular} } &  \multirow{2}{*}{0.19}  & \multirow{2}{*}{0.19} & \multirow{2}{*}{0.18\tnote{b}/0.19\tnote{c}}  \\
                & & &  \\
                
		    \multirow{2}{*}{ \begin{tabular}{c} $\phi_{xy}^-$ \\ $R_5^{-}~(0,a,-a)$ \end{tabular} } & \multirow{2}{*}{0.65} & \multirow{2}{*}{0.67} &  \multirow{2}{*}{0.63\tnote{b}/0.59\tnote{c}}  \\
                & & &  \\
                
                \multirow{2}{*}{ \begin{tabular}{c} $A_R$ \\ $R_4^{-}(0,a,a)$ \end{tabular} }  &  
			\multirow{2}{*}{0.06} & \multirow{2}{*}{0.06} & \multirow{2}{*}{0.06\tnote{b}/0.06\tnote{c}} \\ 
                & & &  \\
                \hline
                \multicolumn{4}{c}{\textbf{\underline{Optical Properties}}}\\
                $\epsilon^{\infty}_{aa}$ &		7.03&	6.02&	  -  \\
                $\epsilon^{\infty}_{bb} $ &		6.52&	5.5&	- \\
                $\epsilon^{\infty}_{xx}$  &		6.77&	5.75&	 $\approx 7.3$\tnote{d,e} \\
                $\epsilon^{\infty}_{cc}$ &		6.15&	5.76&	 $\approx 6$\tnote{d,e} \\
                $E_{Gap}$  &		1.15&	1.77	& 1.1 - 1.9\tnote{f} \\
                \hline     
                \multicolumn{4}{c}{\textbf{\underline{Magnetic Properties}}}\\
                $\mu$ &	3.68	& 3.75 &	3.8 \\
		    $J_{xx}$ = $J_{yy}$ &	-0.59	& -0.25 & -0.83\tnote{b} \\
                $J_{z}$  &  	0.34 &	0.18 &	 0.58\tnote{b} \\
                \multirow{2}{*}{$T_N$} &	\multirow{2}{*}{142} &	\multirow{2}{*}{64} & \multirow{2}{*}{%
                    \begin{tabular}{c}
                        $\sim140$ \\ 
                        (Calc: 207\tnote{b,g})
                    \end{tabular}  
                }	  \\
                & & & \\                
            \end{tabular*}
            \begin{tablenotes}
                \footnotesize
                \item[a] Normalized with respect to the reference phase (Cubic $Pm\overline{3}m$).
                \item[b] Ref. [\onlinecite{Moussa1996}]
                \item[c] Ref. [\onlinecite{Elemans1971238}]
                \item[d] Ref. [\onlinecite{Kovaleva2010}] 
                \item[e] $\epsilon^{\infty}_{xx}$ and $\epsilon^{\infty}_{bb}$ correspond to $\epsilon_{1b}$  and $\epsilon_{1c}$ in the lower frequency range below the first optical transition  in \onlinecite{Kovaleva2010}.
                \item[f] Refs. [\onlinecite{Jung1997,Saitoh1995,PhysRevB.48.17006,Jung1998,Krueger2004,Tobe2001,Moussa1996}]
                \item[g] Calculated in Ref.[\onlinecite{Moussa1996}] with a two J mean-field approach using the measured exchange constants.
            \end{tablenotes}			
        \end{ruledtabular}
        \label{tab:2_all_prop}
    \end{threeparttable}		 	
\end{table}   

In the following we refer to calculated physical quantities using the $(U|J)$ parameters of \textit{Mellan et.al}\cite{Mellan2015} as (8\si{\electronvolt}$|$2\si{\electronvolt}) and our new optimized values as (5\si{\electronvolt}$|$1.5\si{\electronvolt}) and compare them to experimental values.
In the top part of Table \ref{tab:2_all_prop}, we report the relaxed amplitudes of all the modes and strains with imposed AFM-A order. Both tested $(U|J)$ combinations deliver similar strain and mode amplitudes in good agreement with the measured values (maximum deviation for $\phi_{xy}^-(R_5^-)$ $\approx 5\%$).

In the center part of Table \ref{tab:2_all_prop}, we compare the Kohn-Sham band gap and the optical dielectric constant $\epsilon^{\infty}$ obtained with the two GGA+U functionals to experimental data. Both calculated band gaps lie well in the range of experimentally measured values\cite{Jung1997,Saitoh1995,PhysRevB.48.17006,Jung1998,Krueger2004,Tobe2001,Moussa1996}.

The optical dielectric tensor gives a second good measure besides the band gap to test the calculated electronic density.  Refs [\onlinecite{Kovaleva2004,Kovaleva2010}] provide directionally resolved measurements of the optical dielectric tensor at low temperature along the \textit{Pbnm}-c axis and the pseudocubic x-direction to compare with our calculations ($\approx 45 \si{\degree}$ to the orthorhombic a - and b - directions). In Table \ref{tab:2_all_prop} we report the dielectric tensor in the orthorhombic axis as well as rotated to the same crystallographic orientation as in  \cite{Kovaleva2004,Kovaleva2010}, where $\epsilon_{xx}^\infty=\epsilon_{yy}^\infty$, while in the orthorhombic coordinate systems it holds $\epsilon_{aa}^\infty\neq\epsilon_{bb}^\infty $. In the pseudocubic \textit{x,y,z}-system \textit{x} and \textit{y} are not orthogonal, for which reason the off diagonal element $\epsilon_{xy}^\infty\neq 0 $. However, since $\epsilon_{xy}^\infty$ is one magnitude smaller ($<0.5$) than the diagonal terms and as it has not been reported in experiments, we did not note it in Table \ref{tab:2_all_prop}.
 
PBEsol + ($8\si{\electronvolt}\lvert2\si{\electronvolt}$) and PBEsol + ($5\si{\electronvolt}\lvert1.5\si{\electronvolt}$) yield electronic band gaps, which lie well in the range of the experimentally measured ones, although increasing with U.
Regarding the optic dielectric constant, PBEsol + ($5\si{\electronvolt}\lvert1.5\si{\electronvolt}$) yields values in better agreement with experiment, which also reproduce the optical anisotropy absent with PBEsol + ($8\si{\electronvolt}\lvert2\si{\electronvolt}$). 

In the bottom part of Table \ref{tab:2_all_prop} we compare the calculated magnetic properties with experimental values. We made a two \textit{J} exchange constant mean field model, which is sufficient to justify the AFM-A order and can be found in several publications in recent literature \cite{Moussa1996,Mellan2015,Munoz2004}. In our definition, a negative value of \textit{J} indicates ferromagnetic exchange. To calculate the exchange constants, we used the energy differences of the relaxed AFM-A, AFM-G and FM phases and assumed a total spin of 4$\mu_{B}$ per Mn site. In this way the calculated exchange constants contain implicitly spin-phonon coupling, which is however small as the relaxed phases for the different magnetic order are structurally close. Our experimental reference is [\onlinecite{Moussa1996}], where the magnetic exchange constants were derived from magnon dispersion measurements. It is noteworthy, that $T_{N}$ calculated in the mean filed model with the measured exchange constants lies 67 \si{\kelvin} above the measured $T_{N}$ because of the neglect of spin-fluctuations. PBEsol + ($8\si{\electronvolt}\lvert2\si{\electronvolt}$) underestimates both exchange constants by an approximate factor of three. In contrast PBEsol + ($5\si{\electronvolt}\lvert1.5\si{\electronvolt}$) underestimates less the exchange constants with respect to the experiment and finds a Neel-Temperature from mean field theory comparable to the experimental one.

In conclusion,  both ($5\si{\electronvolt}\lvert1.5\si{\electronvolt}$) and ($8\si{\electronvolt}\lvert2\si{\electronvolt}$) produce a good description of the structural ground state of LaMnO$_3$. Considering additionally electronic, optical and magnetic properties, ($5\si{\electronvolt}\lvert1.5\si{\electronvolt}$) provides the better \textit{global} estimate and will be further used in this work. 
        

\section{Potential Energy Surfaces}
\label{sec:VI_PESJTD}

In this section we discuss the shape of the Born-Oppenheimer potential energy surface (PES) around the cubic phase with respect to the key Jahn-Teller distortion in LaMnO$_3$, $Q_{2z}^{\mathbf{M}}$ (See Table \ref{Tab:Define_Qi} and \ref{tab:2_all_prop}).  We quantify mode-mode, mode-strain couplings, and  vibronic Jahn-Teller couplings by successively adding one by one the main lattice distortions found in the \textit{Pbnm} ground state. To do so, we fit the free energy surface by potentials of the type: 

\begin{equation} 
\mathscr{F} = E_0 + \alpha_{JT}  \abs{Q_{2z}^{\mathbf{M}}} + \alpha Q_{2z}^{\mathbf{M}} + \beta (Q_{2z}^{\mathbf{M}})^2 + \gamma (Q_{2z}^{\mathbf{M}})^4, 
\label{eq:first_polynom}
\end{equation}  

where $E_0$ is the energy at $Q_{2z}^\mathbf{M}=0$,  $\alpha_{JT}$ describes the vibronic-coupling terms, $\alpha$ quantifies other linear lattice terms, $\beta$ quadratic lattice terms, and $\gamma$ fourth order terms. In the fit, all modes have been normalized such that 1 corresponds to their ground-state amplitude, which can be found in Table \ref{tab:2_all_prop}.
This approach allows to deduce  \textit{how} the magnetic and structural ground state is reached. The introduction of the absolute function in \eqref{eq:first_polynom} allows to distinquish the vibronic coupling term and linear lattice couplings in the $Q_{2z}^{\mathbf{M}}$ coordinate. The cubic reference lattice parameter is $a_0 \approx 3.935\si{\angstrom}$, which preserves the same volume per formula unit as the bulk ground-state phase. 
The sign and strength of the parameters will be qualitatively discussed in the following sections. A description of the fitting procedure, the whole free-energy expansion, and a Table with the values of the coefficients are given in appendix \ref{sec:Appendix_fit}.

\begin{table}
        \renewcommand{\arraystretch}{1.35}
        \caption{Energy comparison per formula unit of different Magnetic Orderings in the 
         cubic phase of LaMnO$_3$}     
        \begin{tabular*}{\columnwidth}{@{\extracolsep{\fill}}cc}
            \hline \hline
            \hspace{2em} Magnetic Ordering \hspace{2em} &  \hspace{3em} $\Delta E/fu$ (\si{\milli\electronvolt}) \hspace{3em} \\
            \hline
            FM   &  -126.5  \\ 
            AFM-A   &  0.00 \\ 
            AFM-C   &  +175.5  \\ 
            AFM-G   &  +367.9 \\
            \hline \hline      
        \end{tabular*}
        \label{tab:cubic_MO_comp}
\end{table}

\begin{figure}
	\begin{center}
	\includegraphics[width=0.8\columnwidth]{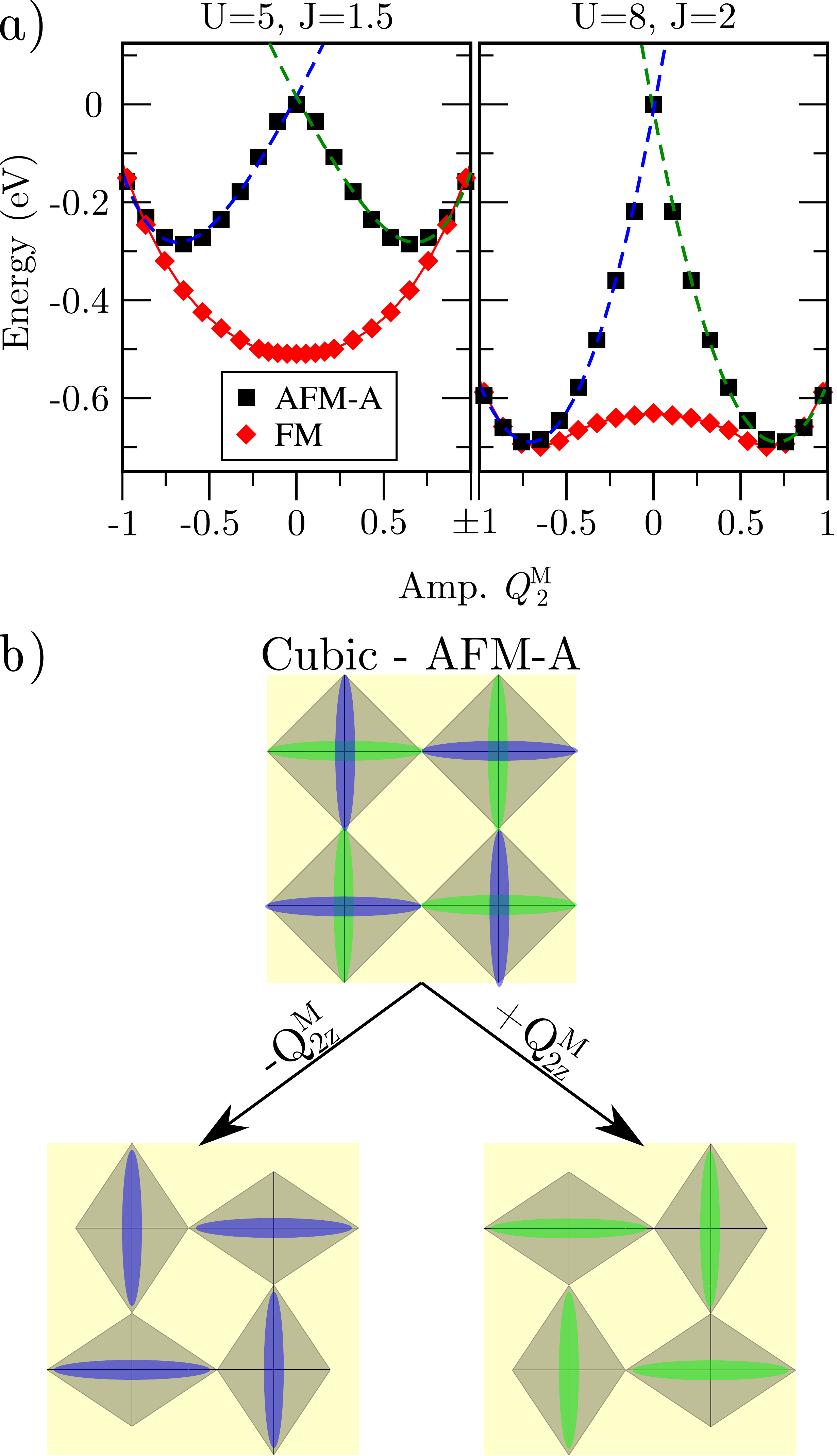}
    \end{center}
	\caption{a) Comparison of the PES of the $Q_{2z}^{\mathbf{M}}$ Jahn-Teller Distortion for different DFT calculation methods used throughout this publication. b) Schematic illustration of orbital orderings, which are degenerate in the cubic structure with AFM-A ordering leading to a metallic phase. A condensation of a $Q_{2z}^\mathbf{M}$ distortion with positive or negative amplitude will stabilize one or the other state. Green and blue colors refer to the dashed lines in Fig. \ref{fig3:Exc_comp}a.} 
	\label{fig3:Exc_comp}
\end{figure}

\subsection{$\mathbf{Q_{2\alpha}^{\mathbf{M}}}$ PES in the cubic phase}
\label{subsec:PES_Q2zM}

 In this section we analyze the relative stability of different magnetic orderings and the stability of $Q_{2z}^{\mathbf{M}}$ distortion in the cubic phase. Inspecting the $Q_{2z}^{\mathbf{M}}$ coordinate is a random choice at this point. Due to the cubic symmetry, the following results would be exactly the same for  $Q_{2x}^{\mathbf{M}}$ and $Q_{2y}^{\mathbf{M}}$. Following KK-approach\cite{ki1982jahn}, we expect an AFM-A magnetic and orbital ordered insulating ground-state with an instability of $Q_{2z}^{\mathbf{M}}$. Following the C-JTE approach we expect an instability of $Q_{2z}^{\mathbf{M}}$ independent of the magnetic order. 
 
 Table \ref{tab:cubic_MO_comp} shows the energy differences per formula unit for different simple magnetic orderings in the cubic phase of LaMnO$_3$. Here our calculations show that the FM ordering is by far the ground state and that large energy differences exists between the different magnetic orders. 
 Fig. \ref{fig3:Exc_comp}a shows the PES of the $Q_{2z}^{\mathbf{M}}$ mode around the cubic $Pm\overline{3}m$ phase and its dependence in terms of the $(U|J)$ parameters. The energy of the cubic AFM-A structure has been set to zero. The amplitude of the $Q_{2z}^{\mathbf{M}}$ distortion has been normalized to the bulk ground-state value. While the differences of the relaxed bulk ground state with respect to the $(U|J)$ parameters are subtle (shown in section \ref{sec:IV_GS}), the differences in Fig. \ref{fig3:Exc_comp}a are rather significant. On the FM surface the $Q_{2z}^{\mathbf{M}}$ distortions changes its character from dynamically stable to unstable for higher U and J values. Similarly, on the AFM-A surface the energy gain of the $Q_{2z}^{\mathbf{M}}$ distortion with respect to the cubic structure is more than twice larger for the larger U and J values. At the opposite, the ferromagnetic ground state and the finite value of $\alpha_{JT}$ on the AFM-A surface are independent of $(U|J)$.
 Fig. \ref{fig3:Exc_comp}a shows that the extraction of \textit{quantitative} parameters from DFT calculations is a difficult task as the numerical value can significantly change with the DFT-approach, while the relaxed ground-state structure might be very similar. However, our results are \textit{qualitatively} the same as the ones of a recent study using a U-value of 3.5 $\si{\electronvolt}$\cite{varignon2019origin}. In appendix \ref{sec:AppendixII_UJ} we show furthermore that the qualitative features shown here with $(U|J)$ = ($5\si{\electronvolt}\lvert1.5\si{\electronvolt.}$)  do not change when applying $(U|J)$ = ($8\si{\electronvolt}\lvert2\si{\electronvolt.}$).   

 The AFM-C and AFM-G surfaces are significantly higher in energy and not shown here, but they also exhibit a vibronic coupling, which is even stronger than in AFM-A.
 
\begin{figure*}
    \centering
      \includegraphics[width=\textwidth]{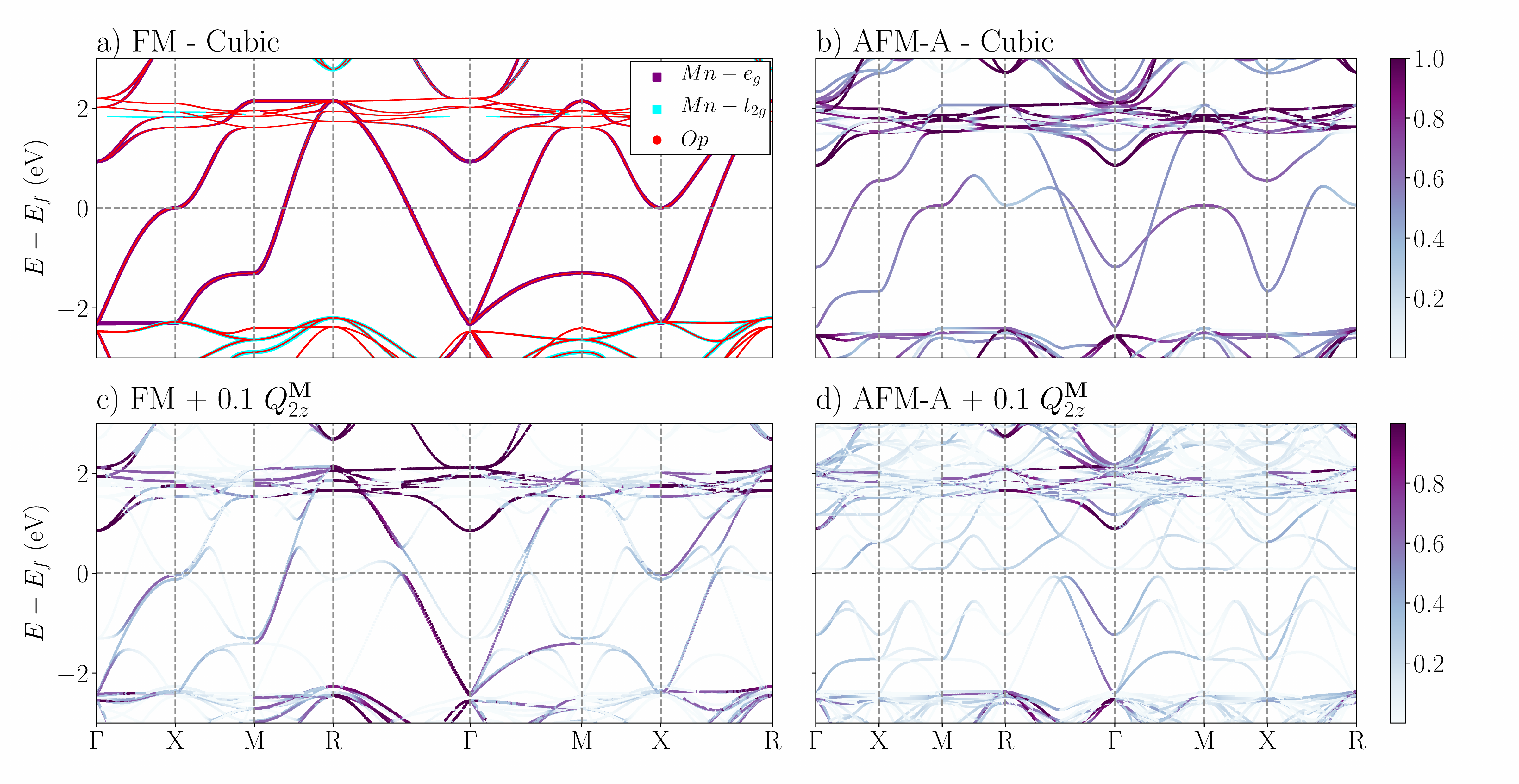}\hfill
    \caption{Electronic Band Structures of LaMnO$_3$ in the range of $\pm 3\si{\electronvolt}$. a) Projection of electronic bands onto Mn-$e_g$, Mn-$t_{2g}$, and O-p orbitals in FM-cubic phase. The size of the dots indicate the character of the bands. b-d) unfolded band structure to cubic Brillouin-zone. The color of the lines indicate the overlap between the supercell and primitive cell k-point. b) AFM-A ordering with cubic atomic positions. c) FM ordering with 10\% $Q_{2\alpha}^\mathbf{M}$ distortion. d) AFM-A ordering with 10\% $Q_{2\alpha}^\mathbf{M}$ distortion, where $\alpha$ is one the cubic lattice directions. In the FM cases the majority spin is shown. In the AFM-A cases one of the two equivalent spin channels are shown.}
    \label{fig4:PROJ-BANDS}
\end{figure*}
 To rationalize the shape of the PES, we can inspect the electronic band-structure (Fig. \ref{fig4:PROJ-BANDS}) in the reference cubic and a distorted structure including 10\% $Q_{2z}^\mathbf{\mathbf{M}}$ (with respect to the ground-state amplitude) distortion in both the FM and AFM-A magnetic orderings. The band structures are unfolded to the cubic Brillouin zone for easy comparison.
 Fig \ref{fig4:PROJ-BANDS}a shows the projection of the band-structures in the cubic phase with FM ordering onto $Mn-e_g$, $Mn-t_{2g}$, and $O-p$ states. In line with other works\cite{Ederer2007,Ku2010,Kovafmmodeheckclseciik2010,Kovafmmodeheckclseciik2011,Kovafmmodeheckclseciik2016}, the band-structure shows that the $e_g$ states are dispersed symmetrically around the Fermi-level, $E_F$, in a range of about $\pm 2 \si{\electronvolt}$. $E_F$ is crossed at the points $X$ and halfway along $M$ - $R$, $\Gamma$ - $R$, $\Gamma$ - $M$ and $X$ - $R$.
 
 If the AFM-A magnetic ordering is imposed (see Fig. \ref{fig4:PROJ-BANDS}b), the local degeneracy at $\Gamma$ of the $e_g$ bands is lifted, showing the symmetry breaking produced by the magnetic order. $E_F$ crosses the $e_g$ bands at $M$ and halfway along $\Gamma$ - $X$, $\Gamma$ - $M$, $\Gamma$ - $R$, $X$ - $M$ and $X$ - $R$. The increase of many of the occupied valence states in the AFM-A cubic case with respect to the FM ordering (e.g. compare the section from $\Gamma$ over $M$ to $X$ of Fig.  \ref{fig4:PROJ-BANDS}a and b) leads to the large increase of the total-energy from FM to AFM-A in the cubic phase (See Table \ref{tab:cubic_MO_comp} and Fig. \ref{fig3:Exc_comp}). The metallicity of the AFM-A cubic phase, despite the local non-degeneracy of the $e_g$ states, can be explained by the degeneracy of two types of orbital orderings within this phase, as schematically drawn in Fig. \ref{fig3:Exc_comp}b. 
  
 If the $Q_{2z}^{\mathbf{M}}$ distortion is added, the electronic bands are split  halfway along all the high-symmetry points (Compare Fig. \ref{fig4:PROJ-BANDS}c and d). The system will gain electronic energy if the $e_g$ bands are crossing the Fermi level at these points as virtual states are shifted to higher energies and occupied ones to lower energies. Moreover, an insulating state can only be created by the application of the $Q_{2z}^{\mathbf{M}}$ distortion if the $e_g$ bands cross the Fermi-level at all the splitting points.
 
 In the FM case only four splitting-points and crossings with the Fermi-level coincide: At $X$ and halfway between  $\Gamma$ - $M$, $\Gamma$ - $R$, $X$ - $R$, and $M$ - $R$. At the other splitting points halfway between $\Gamma$ -  $X$, and $X$ - $M$ the $e_g$ bands are deep in the valence states at about -1.5 \si{\electronvolt} (or one quarter of the $e_g$ bandwidth). The absence of the vibronic coupling can then be explained by 
 \begin{equation}
 \alpha_{JT} = \left . \int_{BZ}\sum_{n=1}^{n_{e^-}}\frac{\partial E_n(\vec{k})}{\partial Q_{2z}^\mathbf{M}}\right|_{Q_{2z}^\mathbf{M}=0} = 0, 
 \label{eq:JT-coupling-FM}
 \end{equation}
 where $E_n(\vec{k})$ is the energy of band $n$ at $\vec{k}$ and its derivatives with respect to $Q_{2z}^\mathbf{M}$ are summed over all occupied states, which are the number of electrons contained in the calculation $n_{e^-}$. Eq. \eqref{eq:JT-coupling-FM} means that, overall, for each k-point at which the total electronic energy is decreased by a variation of $Q_{2z}^\mathbf{M}$ there is another one at which it is increased by the same amount. 
 Finally, in the FM case there is one direction that is unaffected by the $Q_{2z}^\mathbf{M}$ distortion, which can be identified by one band that follows the original $e_g$ paths. Most clearly to be seen at the start of the path from $\Gamma$ over $X$ to $M$ (Compare Fig. \ref{fig4:PROJ-BANDS}a and c). This band accounts for the z-direction in real-space that is not affected by the $Q_{2z}^\mathbf{M}$ distortion.
 
 In the AFM-A case the points at which the condensation of the $Q_{2z}^\mathbf{M}$ distortion splits the $e_g$ bands and their crossing with $E_F$ in the cubic Brillouin zone coincide, such that the $Q_{2z}^\mathbf{M}$ distortion leads to a lowering of the electronic energy and Eq. \eqref{eq:JT-coupling-FM} becomes non zero. Hence the origin of the finite vibronic coupling is a Peierls-like effect where the destruction of the translational symmetry leads to an energy gain. The doubling of the periodicity can be seen most clearly in the oscillations from $\Gamma$ to $X$ and $M$ to $R$. Here magnetic order and $Q_{2z}^\mathbf{M}$ distortion work together in an intriguing way to result in a finite vibronic coupling. Our result shows that future works should focus on the generalization of the spin-structural Peierls-effect in corner shared octahedra networks.
In real space, the condensation of $Q_{2z}^\mathbf{M}$ with positive or negative amplitude corresponds to the stabilization of one orbital order, which will represent an non-degenerate electronic ground state in the distorted phase (See Fig.  \ref{fig4:PROJ-BANDS}d) and Fig. \ref{fig3:Exc_comp}b). The combination of spin and $Q_{2z}^{\mathbf{M}}$ distortion corresponds to the doubling of the periodicity in the three space directions, which stabilizes one distinct orbital order. In the cubic phase both orbital orders are degenerate and explain the metallicity. 
    
Finally, we want to summarize the major result of this section.
    
    (i) The origin of the vibronic coupling on the AFM-A surface appears to be a Peierls-like effect, where AFM-A order and $Q_{2z}^\mathbf{M}$ distortion combine to break the translational symmetry. The conceptual similarity of Peierls effect, CJTE and also KK induced orbital-order has already been noted by \textit{Polinger}\cite{Polinger2007},  \textit{Polinger} and \textit{Bersuker}\cite{bersuker2012vibronic}, and recently by \textit{Streltsov} and \textit{Khomskii} \cite{Streltsov2017}. 
    (ii) The absence of spontaneous orbital-order and the FM ground state in the cubic phase hint that the KK mechanism might not apply to the cubic phase of LaMnO$_3$ and that a renormalization of the intersite electronic parameters might be key to activate it.  
\begin{figure*}[t]
    \includegraphics[width=\textwidth]{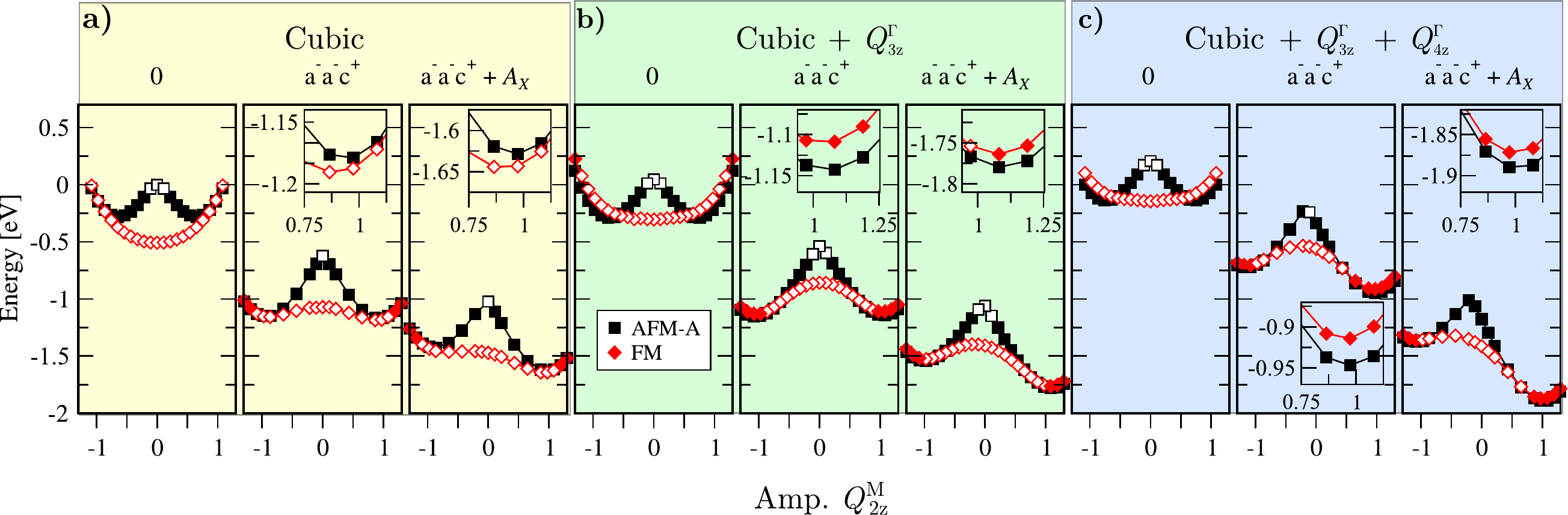}
	\caption{Comparison of the PESs of the $Q_{2z}^{\mathbf{M}}$ mode within different distorted structures. The three main panels refer to distinct unit cells : a) cubic lattice constants $a=b=c=3.935\si{\angstrom}$. b) cubic lattice constant and added tetragonal strain $Q_{3z}^{\mathbf{\Gamma}}$ (as in the ground state). c) cubic lattice constant and added tetragonal $Q_{3z}^{\mathbf{\Gamma}}$ and shear  $Q_{4z}^\mathbf{\Gamma}$ strains, leading to the ground state orthorhombic lattice constants. Within a),b) and c), the three sub-panels refer from left to right to the condensation of $Q_{2z}^{\mathbf{M}}$ mode (i) alone (0), (ii) in presence of octahedral rotations condensed with their ground state amplitudes ($a^-a^-c^+$) and (iii) in presence of  octahedral rotations and antipolar motion $A_X$ condensed with their ground state amplitudes ($a^-a^-c^+ + A_X$). All energies refer to the cubic $Pm\overline{3}m$ structure with AFM-A magnetic ordering, which is set to zero. Open (resp. filled) symbols denote metallic (resp. insulating) states.}
    \label{fig4:Cub_Tet_Orth_comp}
\end{figure*}

\subsection{$\mathbf{Q_{2z}^{\mathbf{M}}}$ PES in presence of other lattice distortions}

In order to investigate under which structural conditions the AFM-A magnetic order is stabilized, we condensed the principal lattice distortions and strains, and sampled the PES in terms of $Q_{2z}^{\mathbf{M}}$ on top of these already distorted structures. The results are shown in Fig. \ref{fig4:Cub_Tet_Orth_comp}.

In Fig. \ref{fig4:Cub_Tet_Orth_comp}a we used the cubic lattice constant $a_0 = 3.935\si{\angstrom}$ and sampled the PES surface along the $Q_{2z}^{\mathbf{M}}$ line successively when appearing (i) alone, (ii) on top of the octahedral rotations $\phi_z^+$ and $\phi_{xy}^-$ with the bulk ground state amplitude, and finally (iii) on top of the rotations plus the $A_X$ motion with their corresponding ground state amplitudes. In Fig. \ref{fig4:Cub_Tet_Orth_comp}b we followed the same procedure for $\phi_z^+$ and $\phi_{xy}^-$ rotations and $A_X$ distortion in the three subpanels, but condensed on top the tetragonal strain $Q_{3z}^{\mathbf{\Gamma}}$ which leads to lattice constants of $a=b=5.66\si{\angstrom}$ and $c=7.61\si{\angstrom}$. Finally in \ref {fig4:Cub_Tet_Orth_comp}c we additionally condensed the shear strain $Q_{4z}^{\mathbf{\Gamma}}$. It leads together with $Q_{3z}^{\mathbf{\Gamma}}$ to the ground state orthorhombic lattice constants. Energies in all graphs are expressed with respect to the same energy reference (cubic AFM-A), allowing the reader to easily find the global ground state under certain conditions.

Calculations are reported for FM and AFM-A orders. Additionally, we note in Fig. \ref{fig4:Cub_Tet_Orth_comp} if the relaxed electronic wave function represents a metallic (open symbols) or insulating state (filled symbols). In this section we limit ourselves to a qualitative discussion of the interplay of lattice and electronic band-structure, without an explicit demonstration of unfolded band-structures. The complete set of fitted coefficients is reported in Appendix \ref{sec:Appendix_fit}.

\subsubsection{Cubic unit cell}
\label{subsec:PES_rot+AX}
Let us first focus on Fig. \ref{fig4:Cub_Tet_Orth_comp}a. The left panel corresponds to the pure cubic lattice and hence to the left panel in Fig \ref{fig3:Exc_comp}a.
Going from no rotations (left panel) to the structure with rotations (middle panel) in the cubic lattice, the global energy is lowered since the rotations are unstable ($E_0^\phi<0$ in Table \ref{tab:Constant_fit}) . Moreover, $Q_{2z}^{\mathbf{M}}$ changes from dynamically stable to unstable on the FM surface. Also, the shifted single wells of the AFM-A surface becomes deeper. 

For the FM-surface this behavior can be attributed to bi-quadratic couplings terms in the free energy expansion between the rotations and the $Q_{2z}^{\mathbf{M}}$ mode 
\begin{equation} 
\mathscr{F} \propto \beta_2 (\phi)^2 (Q_{2z}^{\mathbf{M}})^2, 
\label{eq:rot_Q21}
\end{equation}   
where the coupling constant $\beta_2$ is largely negative and $\phi$ represents a global rotation amplitude that implies that $\phi_{z}^+$ and $\phi_{xy}^-$ keep the same ratio as in the ground-state (see appendix \ref{sec:Appendix_fit}). 

For the AFM-A surface, $\beta_2$ is close to zero. The shift of the single wells has to be attributed to a strong enhancement of the vibronic coupling $\alpha_{JT}$ expressed by the parameter $\lambda_\phi<0$ in Table \ref{tab:Constant_fit} in the Appendix. 
Nonetheless, the ground state is FM and metallic until the largest amplitudes. On the AFM-A surface a band gap opens instantaneously by applying $Q_{2z}^{\mathbf{M}}$.  Both effects ($\lambda_\phi < 0 $ and $\beta_2 < 0$) should be attributed to the strong reduction of the $e_g$ bandwidth (from about 4\si{\electronvolt} to 3 \si{\electronvolt} - not shown here). 

On both magnetic surfaces, the rotations alone induce  a $Q_{2z}^{\mathbf{M}}$ amplitude close to the experimental one. We emphasize that this strong coupling is related to the specific electronic structure of LaMnO$_3$, as other $Pbnm$ perovskites with significant octahedral rotations show only negligible $Q_{2z}^{\mathbf{M}}$ amplitudes (e.g. CaMnO$_3$\cite{Poeppelmeier1982}).
Additionally to Eq. \ref{eq:rot_Q21}, there is a fourth-order term coupling linearly the $Q_{2z}^{\mathbf{M}}$ mode with the rotations: 
\begin{equation} 
 \mathscr{F} \propto \alpha_1  [(\phi_{xy}^-)^2\phi_{z}^+]Q_{2z}^{\mathbf{M}}.
 \label{eq:rot_Q22}
\end{equation} 
The effect of this term appears however negligible since the symmetry of the potential well is (almost completely) maintained when the rotations are condensed.

Although $A_X$ is intrinsically stable in the cubic phase, its presence is naturally driven in presence of $\phi_{xy}^{-}$ and $\phi_z^+$ due to a trilinear coupling term, 
\begin{equation}
 \mathscr{F} \propto \alpha (\phi_{xy}^{-} \phi_z^{+}) A_X,
\label{eq:trilin_HIF}
\end{equation}
which, in another context, is known to be related to the appearance of hybrid improper ferroelectricity\cite{Benedek2015} in some cation ordered perovskite superlattices. Consequently, as in other $Pnma$ perovskites \cite{Miao-2013}, the condensation of $A_x$ globally decreases the energy through this term (part of $E_0^{\phi,A_X}$ in Eq. \ref{Eq-A4}), producing a rigid down-shift of the energy wells in Fig. \ref{fig4:Cub_Tet_Orth_comp}.

In a similar way there is a trilinear term 
\begin{equation} 
\mathscr{F}\propto\alpha_2 (A_X\phi_{xy}^-)Q_{2z}^{\mathbf{M}}.
\label{eq:rot_X_Q2}
\end{equation}
This term does significantly break the symmetry of the $Q_{2z}^{\mathbf{M}}$ surface on the contrary to term in Eq. \eqref{eq:rot_Q22}. The asymmetry created by the crystal field, induced by the combination of $\phi_{xy}^-$ and $A_X$, is independent of the magnetic order since the fitted coefficient $\alpha_2$ takes close values for AFM-A and FM ordering (See Table \ref{tab:Constant_fit}). That being said, the ground state surface is FM for all structures with cubic lattice constants. Only the AFM-A surface shows insulating behavior around its minima. The coupling terms above are obviously equally valid in the strain distorted unit-cells and similar trends can be seen in the energy surfaces of all three examined cases.  

To summarize, we have found here (i) that octahedral rotations trigger $Q_{2z}^{\mathbf{M}}$ by $e_g$ bandwidth reduction and (ii) that octahdral rotations combined with the antipolar motion $A_X$ induce a significant crystal field that breaks the symmetry of the $Q_{2z}^{\mathbf{M}}$ PES.

\subsubsection{Tetragonally compressed unit cell}

Let us now focus on Fig. \ref{fig4:Cub_Tet_Orth_comp}b in which the compressive tetragonal strain $Q_{3z}^{\mathbf{\Gamma}}$ has been added to the cubic lattice. Again, the PES in terms of $Q_{2z}^{\mathbf{M}}$ is shown when condensing or not the other lattice distortions. Adding $Q_{3z}^{\mathbf{\Gamma}}$ increases energy independently of the magnetic order, but decreases their distance at $Q_{2z}^\mathbf{M}=0$ as
\begin{equation}
 E_0^{Q_3^\mathbf{\Gamma}}({\rm AFM-A})<E_0^{Q_3^\mathbf{\Gamma}}({\rm FM}).
 \end{equation}
 
 On the FM surface, the $Q_{2z}^{\mathbf{M}}$ mode gets significantly softer, compared to panel (a). The softening can be associated to linear-quadratic and bi-quadratic strain-phonon coupling terms 
\begin{equation} 
\mathscr{F} \propto \beta_4Q_{3z}^{\mathbf{\Gamma}}(Q_{2z}^{\mathbf{M}})^2+\beta_5(Q_{3z}^{\mathbf{\Gamma}})^2(Q_{2z}^{\mathbf{M}})^2.
\label{eq:tet-strain-phon}
\end{equation} 
Here the linear-quadratic term is dominating since $\beta_4>\beta_5$ (see Appendix). This implies directly that the appearance of  $Q_{2z}^{\mathbf{M}^{2}}$ favors a compressive over a elongating tetragonal strain $Q_{3z}^{\mathbf{\Gamma}}$ and vice versa. 

On the AFM-A surface it is mainly the electronic instability $\alpha_{JT}$ that is affected by $Q_{3z}^{\mathbf{\Gamma}}$ ( $\lambda_{Q_{3z}^{\mathbf{\Gamma}}}<0$, see Appendix) and shifts the amplitude of $Q_{2z}^{\mathbf{M}}$ close to the experimental bulk value. 
Most interestingly, the ground state surface is no longer the FM one: if the $Q_{3z}^{\mathbf{\Gamma}}$ strain and $Q_{2z}^{\mathbf{M}}$ distortion are condensed  together, the AFM-A energy becomes lower than the FM energy at about $100\%~Q_{3z}^{\mathbf{\Gamma}} + 50\%~Q_{2z}^{\mathbf{M}}$. The linear-quadratic and bi-quadratic strain-phonon coupling terms do exist between the tetragonal strain and all symmetry adapted modes condensed in the \textit{Pbnm} phase. 

Octahedral rotations ($\phi$) and $Q_{3z}^{\mathbf{\Gamma}}$ shift the energy minima on both magnetic surfaces to values well above 1, which can be explained by the phonon-phonon couplings highlighted in Eq. \eqref{eq:rot_Q21}-\eqref{eq:rot_X_Q2}. Nonetheless, the ''cubic plus rotations'' surfaces stay lower in energy than tetragonal strained ones. Interestingly, at this point, the minima on the FM surface become insulating states. We can attribute this to the combined translational symmetry breaking of the anti-phase rotation $\phi_{xy}^-$ and the tetragonal compression of $Q_{3z}^{\mathbf{\Gamma}}$, which together break the translational symmetry just like the AFM-A order. 

Adding $A_X$ breaks the symmetry of the energy surface. The difference of energy between the minima along the positive and negative paths of $Q_{2z}^{\mathbf{M}}$  is increased, due to a quadri-linear strain-phonon term, 
\begin{equation} 
\mathscr{F} \propto \alpha_3 (Q_{3z}^{\mathbf{\Gamma}}\phi_{xy}^{-}A_{X})Q_{2z}^\mathbf{M}.
\label{eq:tet-strain-phon2}
\end{equation} 
We note that the same term exists replacing  $Q_{2z}^{\mathbf{M}}$ with the in-phase octahedral rotation $\phi_z^+$. It is because of those two terms that eventually the tetragonal phase gets slightly stabilized over the cubic one.

To summarize, we have additionally found here (i) that the tetragonal compressive strain $Q_{3z}^{\mathbf{\Gamma}}$ combined with $Q_{2z}^{\mathbf{M}}$ stabilizes the AFM-A against the FM ordering , and (ii) that $Q_{3z}^{\mathbf{\Gamma}}$ combined with the antiphase rotation $\phi_{xy}^-$ induces the opening of a band gap on the FM surface around the $Q_{2z}^{\mathbf{M}}$ minima.

\subsubsection{Orthorhombic ground sate unit cell}

\begin{figure}
\centering 
\includegraphics[width = 7cm]{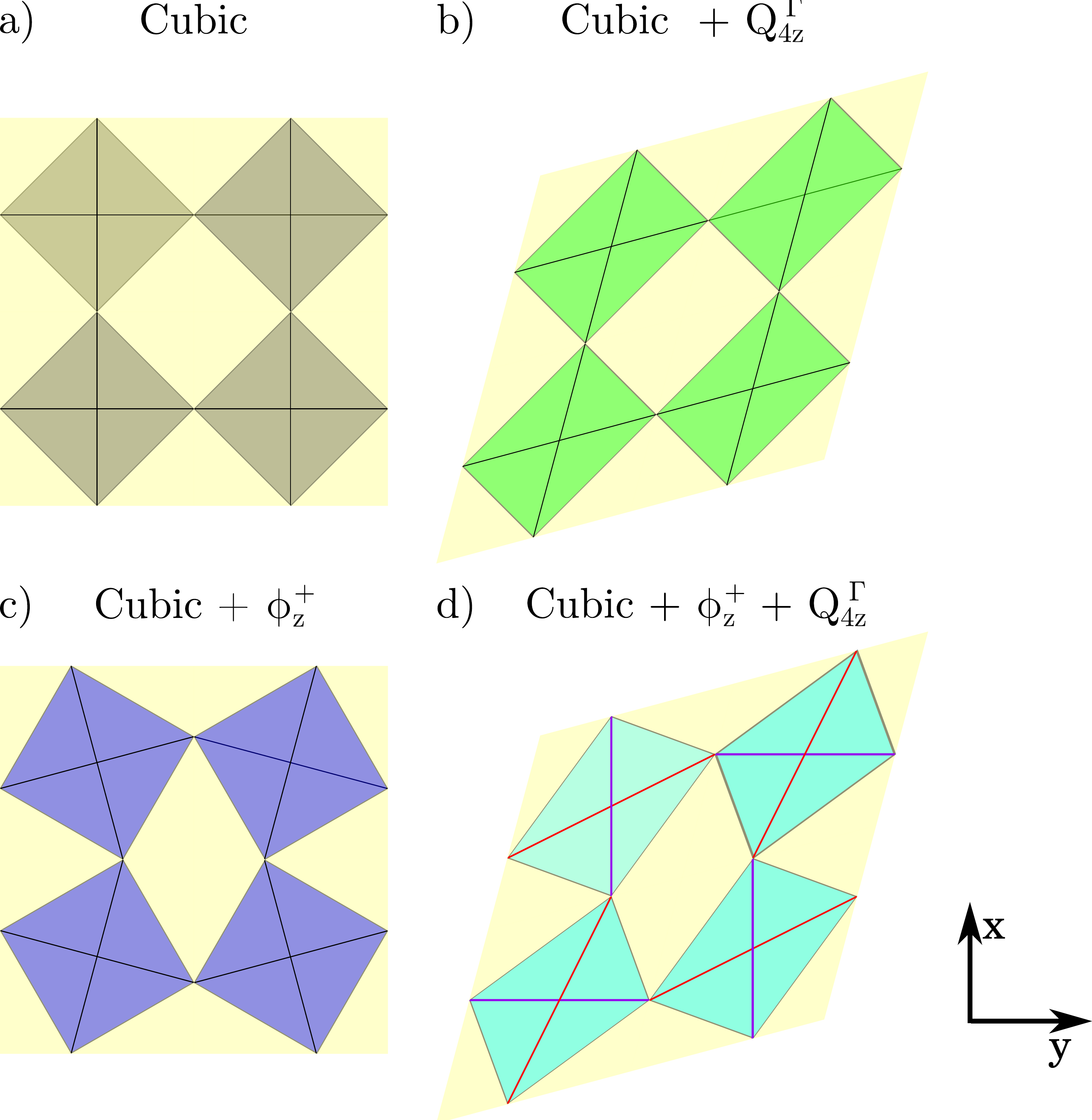}
\caption{Schematic illustration of octahedral rotation $\phi_z^+$  and shear strain $Q_{4z}^\mathbf{\Gamma}$ acting together as a $Q_{2z}^{\mathbf{M}}$ Jahn-Teller distortion of the oxygen octahedra. a) cubic phase, b) phase with shear strain $Q_{4z}^{\mathbf{\Gamma}}$, c) phase with rotation of the octahedra $\phi_z^+$, and d) phase combining shear strain $Q_{4z}^{\mathbf{\Gamma}}$ and rotation $\phi_z^+$. The combination of $Q_{4z}^{\mathbf{\Gamma}}$ and $\phi_z^+$ acts as an effective $Q_{2z}^{\mathbf{M}}$ distortion : while all octahedral axis are equivalent (black lines) in (a), (b) and (c), the combination of $Q_{4z}^{\mathbf{\Gamma}}$ and $\phi_z^+$ in (d) gives rise to shortened (purple) and elongated (red) octahedral axis in a way similar to  $Q_{2z}^{\mathbf{M}}$.}
\label{fig5:Effec_JT}
\end{figure}

Let us finally focus on Fig. \ref{fig4:Cub_Tet_Orth_comp}c, adding together the compressive tetragonal strain $Q_{3z}^{\mathbf{\Gamma}}$ and the orthorhombic shear strain $Q_{4z}^\mathbf{\Gamma}$ with their ground state values in the cubic lattice. The strained unit cell has then the lattice parameter of the relaxed ground state cell. 
Adding the shear strain $Q_{4z}^\mathbf{\Gamma}$ on top of $Q_{3z}^{\mathbf{\Gamma}}$ further increases the global energy, if no other modes are condensed. The distance between the magnetic surfaces is approximately unaltered.
On the contrary to the cubic and tetragonal cases, the symmetry of the $Q_{2z}^{\mathbf{M}}$ PES is broken when octahedral rotations are condensed. This is due to a trilinear term :
\begin{equation} 
\mathscr{F} \propto \alpha_4 (Q_{4z}^{\mathbf{\Gamma}}\phi_z^+)Q_{2z}^{\mathbf{M}}.
\label{eq:shear_rot_q2}
\end{equation} 
Fig. \ref{fig5:Effec_JT} provides a sketch highlighting the physical meaning of Eq. \eqref{eq:shear_rot_q2}. Neither a octahedral rotation nor a shear strain alone can lift the degeneracy of the octahedral bond lenghts. However, when the rotation axis and the axis normal to the shear plane coincide, they act together as an effective $Q_2$ motion  and split the bond lengths. If the rotation is antiphase ($\phi^-$) the effective motion is $Q_{2}^{\mathbf{R}}$, if it is in-phase ($\phi^{+}$) it becomes $Q_{2}^{\mathbf{M}}$. Hence in LaMnO$_3$  $\phi_z^+$ and $Q_{4z}^{\mathbf{\Gamma}}$ build an effective $Q_{2z}^{\mathbf{M}}$ motion. This effective $Q_{2z}^{\mathbf{M}}$ motion explains that once $\phi_z^+$ and $Q_{4z}^{\mathbf{\Gamma}}$ are condensed, the metal to insulator transition is reached for smaller $Q_{2z}^{\mathbf{M}}$ amplitudes compared to the previously discussed surfaces. Finally it also explains, why the gradient discontinuity does not appear at $Q_{2z}^{\mathbf{M}}=0$. To fit the PES with $Q_{4z}^{\mathbf{\Gamma}}$ and $\phi_z^+$ condensed together, we had to introduce a shift of the zero coordinate of $Q_{2z}^{\mathbf{M}}$, which extracts the amplitude of their effective $Q_{2z}^{\mathbf{M}}$ distortion. When  $Q_{4z}^{\mathbf{\Gamma}}$ and $\phi_z^+$ are codensed with their ground-state amplitudes, $Q_{2z}^\mathbf{M}$ takes already $\approxeq 15\%$ of its ground-state amplitude (i.e.  $0.06\si{\angstrom}$). It can be extracted in Fig. \ref{fig4:Cub_Tet_Orth_comp}c at the position of the gradient discontinuity on the AFM-A surface. Despite the trilinear term in Eq. \eqref{eq:shear_rot_q2}, tetragonally and sheared distorted unit cell stay higher in energy compared to the cubic case if only the octahedral rotations are present. It is eventually $A_X$ that induces a orthorhombic ground state through a quartic term, linear in $Q_{2z}^{\mathbf{M}}$ and similar to Eq. \eqref{eq:tet-strain-phon2} :
\begin{equation} 
	\mathscr{F} = \alpha_5(Q_{4z}^{\mathbf{\Gamma}}A_{X}\phi_{xy}^-)Q_{2z}^{\mathbf{M}}.
\label{eq:shear-strain-phon}
\end{equation} 
The FM surface is also insulating around its $Q_{2z}^{\mathbf{M}}$ minima and the AFM-A surface is the global ground state in all $Q_{3z}^{\mathbf{\Gamma}}~+~Q_{4z}^\mathbf{\Gamma}$ distorted cases.

To summarize, we have found here (i) that $\phi_z^+$ and $Q_{4z}^{\mathbf{\Gamma}}$ act as an effective $Q_{2z}^{\mathbf{M}}$ motion and (ii) a quartic term that stabilizes the ground state unit cell shape.

\subsubsection{Summary}

From the discussion of the PESs, we can reach the following conclusions.

(i) Octahedral rotations trigger $Q_{2z}^{\mathbf{M}}$ by a negative bi-quadractic coupling on the FM surface and by an enhanced vibronic coupling on the AFM-A surface. This is attributed to a reduced $e_g$ bandwidth.
 
(ii) Tetragonal strain $Q_{3z}^{\mathbf{\Gamma}}$ is responsible for the magnetic FM - AFM-A transition, by reducing the energy-difference between the AFM-A and FM surface. We note that this is in line with recent ab-initio studies \cite{Rivero2016,Rivero2016a,Hou2014} and an experimental study of FM LaMnO$_3$ thin films grown on SrTiO$_3$\cite{Roqueta2015}. Here, the canonical Jahn-Teller distortion notations allowed us to extract $Q_{3z}^{\mathbf{\Gamma}}$ as the key structural parameter.

(iii) On the FM-surface, a band gap can only be opened by $Q_{2z}^{\mathbf{M}}$ in the presence of compressive tetragonal strain $Q_{3z}^{\mathbf{\Gamma}}$ and the antiphase rotation $\phi_{xy}^-$. This is assigned to the combined strong symmetry breaking of $Q_{3z}^{\mathbf{\Gamma}}$  and $\phi_{xy}^-$ along the \textit{Pbnm}-c axis equivalent to the symmetry breaking of AFM-A order.

(iv) In none of the tested structures we found a finite value of $\alpha_{JT}$ on the FM surface. There is no vibronic coupling in the FM surface with respect to $Q_{2z}^{\mathbf{M}}$.

(v) Various lattice couplings lead to almost identical ground-state structures for FM and AFM-A orderings. This explains the absence of a structural distortion at the magnetic transition $T_N\approx 140K$.

(vi) Shear strain $Q_{4z}^{\mathbf{\Gamma}}$ and in-phase octahedral rotation $\phi_z^+$ act as an effective $Q_{2z}^{\mathbf{M}}$ distortion.


\section{$\mathbf{Q_{2z}^{\mathbf{M}}}$ and other Lattice Distortions around the $\mathbf{T_{JT}}$ transition}

In this section we analyze the temperature evolution of the amplitudes of all relevant strains and phonon modes around the orbital ordering transition at $T_{JT} \approx 750\si{\kelvin}$, as measured experimentally. We discuss the variation of the amplitudes of lattice modes and strains in connection with the coupling terms defined before. We recalculate the $Q_{2z}^\mathbf{M}$ PES within the measured experimental structures around the transition. We show by a simple Monte-Carlo (MC) simulation of the evolution of $Q_{2z}^\mathbf{M}$ amplitude with temperature that those PESs qualitatively reproduce the phase transition. Our approach highlights that an important contribution to the stabilizing energy of the insulating phase is the spin symmetry breaking, which appears dynamically in the PM phases around the MIT but can be properly extracted from static DFT calculations.

Our experimental source is the recent study of \textit{Thygesen et al.}\cite{Thygesen2017}, in which the authors measured the lattice structure over $T_{JT}$  between 300 \si{\kelvin} and 1000 \si{\kelvin}. The aim of their study was to identify the differences in the local structure of the orbital ordered \textit{O'} and disordered \textit{O} phases in order to derive a better understanding of the \textit{O} phase (sometimes also called orbital-liquid phase and the transition has been described as orbital melting\cite{Chatterji2003,Trokiner2013}).

\begin{figure}
    \includegraphics[width=\columnwidth]{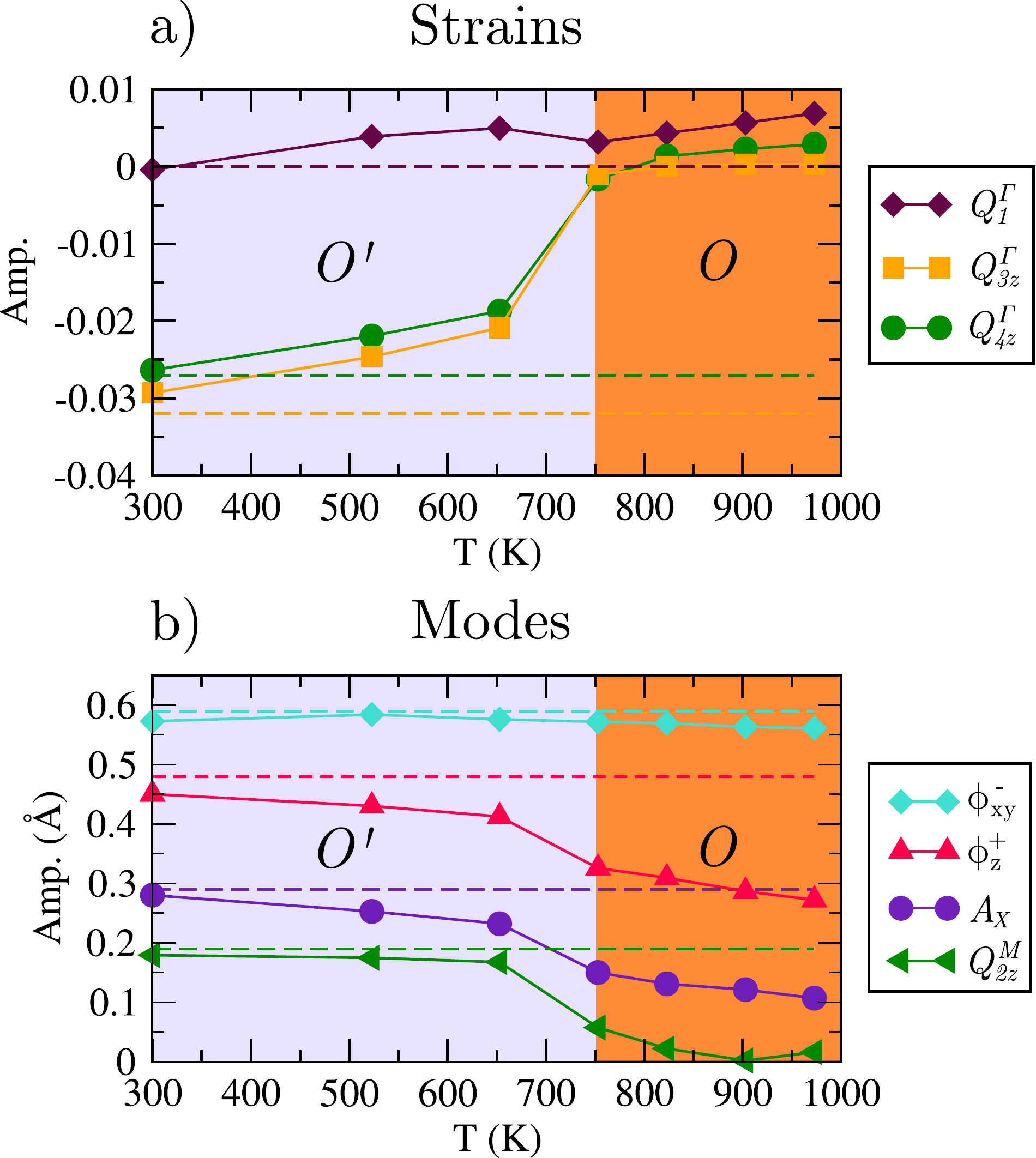}
    \caption{Experimental lattice modes and strain amplitudes across the \textit{O'/O}-transition at $T_{JT}\approx750K$. Structures extracted from  Ref. \onlinecite{Thygesen2017} and analyzed with \textsc{isodistort}. Dashed lines show  low temperature amplitudes.}
    \label{fig:Mode_Strain_Analysis}
\end{figure}

\begin{figure}
    \includegraphics[height=62.5em]{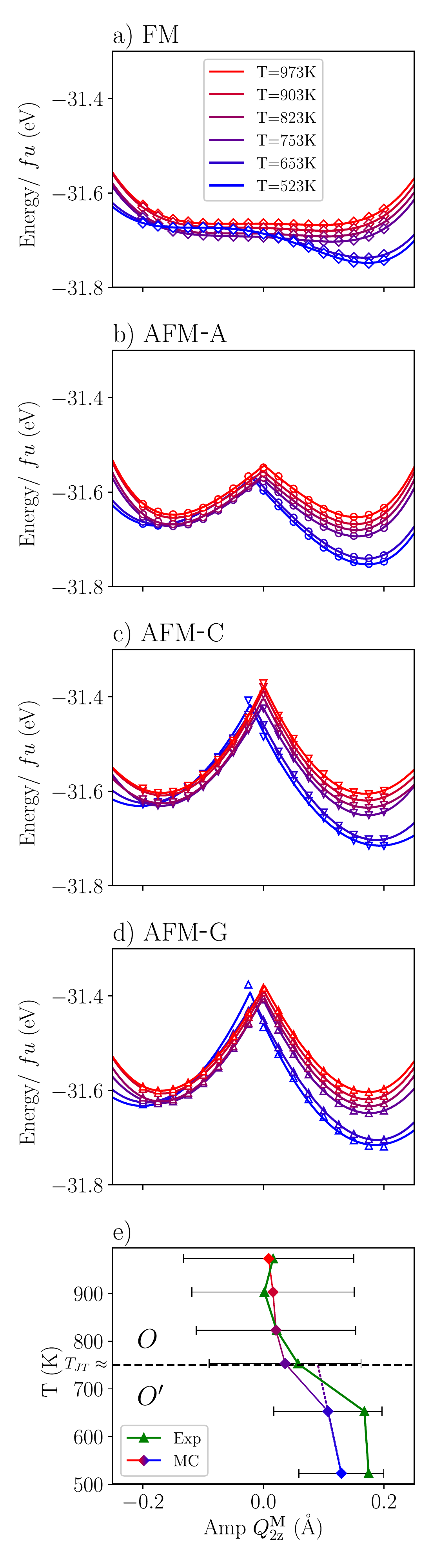}
    \caption{a)-d) PES of $Q_{2z}^{\mathbf{M}}$ distortion as calculated from DFT within the lattice structures measured by \textit{Thygesen et al.} at the indicated temperatures and magnetic orders. Markers show the DFT energies, continuous lines a polynomial fit. e) Experimental amplitudes of $Q_{2z}^{\mathbf{M}}$ and mean amplitudes resulting a Monte-Carlo (MC) sampling of the above PESs with $T_{sim}/T_{exp} = 0.625$. Error Bars show the standard deviation of the MC simulation.}
    \label{fig:MC_Q2M}
\end{figure}
In Fig. \ref{fig:Mode_Strain_Analysis}a we show the symmetry adopted strain and in Fig. \ref{fig:Mode_Strain_Analysis}b the symmetry adapted phonon mode analysis of the experimental data around $T_{JT}$. The low temperature amplitudes noted in Table \ref{tab:2_all_prop} are shown as dashed lines. Additionally, we show the variation of the unit-cell volume through the volume strain $Q_1^{\mathbf{\Gamma}}$, which points out the well-known volume collapse at $T_{JT}$\cite{Maitra2004,Chatterji2003,Ahmed2009,Thygesen2017}. The tetragonal strain $Q_{3z}^{\mathbf{\Gamma}}$ and shear strain $Q_{4z}^{\mathbf{\Gamma}}$ show a linear decrease in amplitude for temperatures lower than $T_{JT}$. At $T_{JT}$ they abruptly disappear almost completely and have only small amplitudes in the orbital disordered \textit{O} phase. From the inspection of symmetry strains in Fig. \ref{fig:Mode_Strain_Analysis}a, it is obvious that the disappearance of $Q_{3z}^{\mathbf{\Gamma}}$ and  $Q_{4z}^{\mathbf{\Gamma}}$ are much more severe at $T_{JT}$ than the volume collapse of $Q_1^{\mathbf{\Gamma}}$. Although this has been  previously pointed out  by \textit{Carpenter and Howard}\cite{Carpenter2009a}, recent studies continue to emphasize the volume collapse\cite{Thygesen2017}. 

The amplitudes of the modes at 300 \si{\kelvin} are close to their low temperature values. The amplitude of the antiphase rotations $\phi_{xy}^-$ stays approximately constant and close to the low temperature value across the whole temperature range from 300 \si{\kelvin} to 1000 \si{\kelvin}. The amplitudes of the in-phase rotation $\phi_{z}^+$ and the antipolar motion $A_X$ decrease linearly between 300 \si{\kelvin} and  $T_{JT}$. The Jahn - Teller distortion $Q_{2z}^{\mathbf{M}}$ keeps an almost constant amplitude between 300 \si{\kelvin} and  $T_{JT}$. At $T_{JT}$ there is a discontinuity for  $\phi_{z}^+$,  $A_X$, and $Q_{2z}^{\mathbf{M}}$ with a sudden reduction in their amplitude. However, $Q_{2z}^{\mathbf{M}}$ does not completely disappear directly at $T_{JT}$ as it could be expected. Above $T_{JT}$, $\phi_{z}^+$,  $A_X$, and $Q_{2z}^{\mathbf{M}}$ continue to decrease linearly ($Q_{2z}^{\mathbf{M}}$ until it reaches approximately zero amplitude at $\approx 900~\si{\kelvin}$).

The similar linear temperature dependence of $\phi_{z}^+$,  $A_X$, $Q_{2z}^{\mathbf{M}}$ in the \textit{O'} and \textit{O} phases can be easily explained by Eq. \eqref{eq:trilin_HIF} and \eqref{eq:rot_X_Q2}. The amplitude change of $\phi_{z}^+$ should be associated as the driving force as $A_X$ is stable by itself and the amplitude of $\phi_{xy}^-$ is nearly constant . Then $A_X$ follows simply the amplitude of $\phi_{z}^+$ through the trilinear coupling in Eq. \eqref{eq:trilin_HIF}. 
Consistently, $Q_{2z}^{\mathbf{M}}$ follows the amplitude of $\phi_{z}^+$ through the trilinear coupling in Eq. \eqref{eq:rot_X_Q2}.

The small but non-zero amplitude of $Q_{2z}^{\mathbf{M}}$ just before the transition might suggest that the variation of $\phi_{z}^+$ with temperature induces the transition by the trilinear \textit{improper} mechanism of Eq. \eqref{eq:rot_X_Q2}. 

To get a more detailed insight, we recalculated the PESs of $Q_{2z}^{\mathbf{M}}$ in the experimental structures extracted from Ref. [\onlinecite{Thygesen2017}] between 523K and 973K and then executed a simple MC simulation on this surfaces to find the mean amplitude of $Q_{2z}^{\mathbf{M}}$ at a given temperature. To account for the PM state at the transition, we calculated the PESs in four distinct simple magnetic orders (FM, AFM-A, AFM-C and AFM-G, see Fig. \ref{fig:MC_Q2M}a-d). Then we execute a MC simulation of the $Q_{2z}^{\mathbf{M}}$ amplitude on each magnetic surface individually. We perform a 100 times 10 million MC steps from which we extract the mean $Q_{2z}^{\mathbf{M}}$ amplitude and the standard deviation of the amplitude. Finally, we find the overall mean amplitude as the mean of the four surfaces. The resulting mean amplitude is shown alongside the measured one in Fig. \ref{fig:MC_Q2M}e. Error bars show the standard deviation of the $Q_{2z}^{\mathbf{M}}$ amplitude, which gives a measure on how much $Q_{2z}^{\mathbf{M}}$ fluctuate at a given temperature. The approach of mixing together the simulation of various magnetic orderings can be seen as a simplified account for the multi Slater-determinant character of a PM electronic wavefunction that is more rigorously treated by advanced material specific many-body methods like DFT+DMFT. On the other hand DFT+DMFT calculations typically exclude the influence of the lattice by parametrizing an Anderson impurity model on a given lattice configuration.

Although a renormalization of the temperature is needed ($T_{sim}/T_{exp}=0.625$), it can be seen that the qualitative features of the $Q_{2z}^{\mathbf{M}}$ amplitude with reducing temperatures are well reproduced by our simple simulation approach: in particular, a small linear increase of $Q_{2z}^{\mathbf{M}}$ before the transition and a sudden jump to larger amplitudes below. The error bars show a huge distribution  above $T_{JT}$, which is consistent with the experimentally described \textit{liquidish} behavior, and a strong reduction of the distribution below. 

Through the PESs, we can examine the origin of this transition. The FM surface shows that the rotation amplitudes of $\phi_{xy}^-$ and $\phi_{z}^+$ are large enough even at the highest temperature to produce a weak instability through the bi-quadratic coupling in Eq. \eqref{eq:rot_Q21}. Then through the trilinear coupling of Eq. \eqref{eq:rot_X_Q2}, a weak asymmetry of the surface is induced which increases before the transition. After the transition this asymmetry is significantly amplified such that he minimum on the negative side of $Q_{2z}^{\mathbf{M}}$ disappears. This change can be mainly attributed to the relaxation of the strains $Q_{3z}^{\mathbf{\Gamma}}$ and  $Q_{4z}^{\mathbf{\Gamma}}$ and the associated couplings in Eq. \eqref{eq:tet-strain-phon},\eqref{eq:tet-strain-phon2},\eqref{eq:shear_rot_q2},\eqref{eq:shear-strain-phon}, which are linear in $Q_{2z}^{\mathbf{M}}$. Only taking into account the FM surface a lattice triggered picture would be convincing. However, the minima on this surface are much too shallow to explain the transition at such a high temperature.

The shallow minima are corrected by taking into account the AFM PESs to mimic the PM phase. On the AFM PESs, deep minima exist due to the Peierls condition that is met in all AFM orderings. This translates into a finite vibronic coupling, whose strength is increased going from AFM-A over AFM-C to AFM-G as the $e_g$ bandwidth is decreased. Taking the AFM PESs into account in the MC-simulation instead of relying only on the FM surface increases the transition temperature strongly. This underlines the importance of \textit{spin symmetry breaking} to activate the strong electron-lattice coupling stabilizing the insulating phase. Also, it shows that the importance of the dynamical fluctuations of the spin in the PM phase for stabilizing the insulating phase lies only partly in the activation of dynamical correlations, but equally in the  \textit{instantaneous} symmetry breaking it produces. The effect of spin symmetry breaking, although experimentally dynamical, can be properly extracted by static DFT calculations as shown from our simplistic approach.

 
Finally, a multifaceted image about the origin of the transition emerges. On one hand it is ``improperly'' induced by the lattice, favoring one side of the $Q_{2z}^{\mathbf{M}}$ well over the other. On the other hand, it incorporates also the characteristics of an order-disorder transition as deep minima for $Q_{2z}^{\mathbf{M}}$ persist in the high temperature \textit{O}-phase, which is magnetically and structurally disordered. The origin of these deep minima is the \textit{dynamic symmetry breaking} of the spin-order in the PM phase. 

However, our  MC approach does not allow us to comment on the persistent debate of the importance of \textit{dynamic correlations} over \textit{dynamic spin symmetry breaking} and lattice symmetry  breaking\cite{Varignon2019}. Nonetheless, the least we can deduce is that \textit{instantaneous electronic symmetry breaking} contributes to a large part of the stabilization energy that drives the MIT by inducing a large electron-phonon coupling. It has been noted before that such a large electron-phonon coupling is necessary to explain the high MIT transition temperatures and the dynamic Jahn-Teller deformations in the high-temperature phase in KCuF$_3$\cite{Pavarini2008} and LaMnO$_3$\cite{Pavarini2010}.

Finally, we note that it would be possible to optimize $(U|J)$ values that bring the MC simulation transition temperature to the experimental transition temperature. This would however not lead to any additional insights as it would merely mean to tune a parameter in a reduced model description. 
To gain more microscopic insight into the transition mechanism and the dynamical properties of the high temperature metallic \textit{O}-phase, nucleic and electronic subsystems have to be treated \textit{dynamically coupled} in large supercells at finite temperature. An approach to realize such a dynamic coupling are so-called second principles models\cite{Wojdel2013,Garcia-Fernandez2016,Escorihuela-Sayalero2017}.



\section{Charge vs. Orbital Ordering in LaMnO$_\mathbf{3}$}


\begin{figure}
    \centering
    \includegraphics[width=0.8\columnwidth]{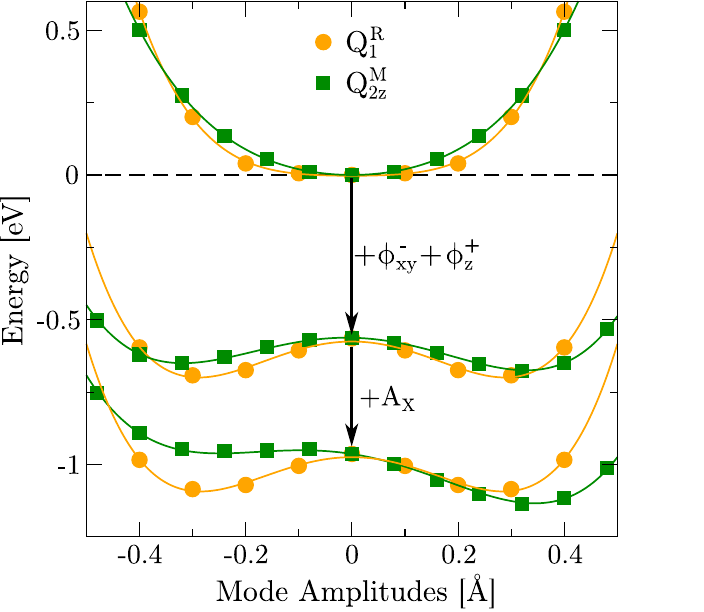}
    \caption{PES of $Q_1^{\mathbf{R}}$ and $Q_{2z}^{\mathbf{M}}$ distortions within FM ordering in cubic-structure (top curves), with condensed octahedral rotations $\phi_{xy}^-$ and $\phi_{z}^+$ - (middle curves), and with additionally condensed antipolar motion $A_X$ (bottom curves).}
    \label{fig:Q1R_vs_Q2M}
\end{figure}

Until this point, we investigated the relevant \textit{statically appearing} distortions in the single-crystal ground state phase of LaMn$O_3$. However, at few occasions, a charge ordering instability has been discussed as an alternative and competing mechanism to orbital ordering\cite{Whangbo2002,Mazin2007} or as the origin of the transition in the high-temperature orbital liquid phase, which has been, in that picture, described as an electron-hole liquid phase\cite{Zhou1999,Moskvin2009}.
Such a charge-ordering instability in the high-temperature phase should be accompanied by the instability of a \textit{breathing type distortion} $Q_1^{\mathbf{R}}$ (see Table \ref{Tab:Define_Qi}). Recent works showed that the charge-ordering transition in RNiO$_3$ (a $e_g^1$ perovskite with doubly occupied $t_{2g}$ states) can be understood as a Peierls transition \cite{Mercy2017} triggered by the appearance of octahedral rotations. Moreover, the same picture applies to alkaline-earth ferrites, AFeO$_3$\cite{Zhang2018}, with the same formal occupation of Fe \textit{d}-states as Mn \textit{d}-states ($d^4$ = $t_{2g}^3 e_g^1$). In those ferrites, the instabilities of $Q_1^{\mathbf{R}}$ and $Q_{2z}^{\mathbf{M}}$ compete and can be tuned by epitaxial strain. A similar behavior has been found in HoNiO$_3$ \cite{Jiang2018}.
 
In Fig. \ref{fig:Q1R_vs_Q2M}, we show that the same competition exists for the RMnO$_3$ series with the example of LaMnO$_3$. Here, we limit ourselves to caclulations within the FM ordering. In the top of Fig. \ref{fig:Q1R_vs_Q2M}, the PESs of $Q_1^{\mathbf{R}}$ and $Q_{2z}^{\mathbf{M}}$ within cubic LaMnO$_3$ can be seen. Both of them show stable single wells, with comparable harmonic and higher-order dependencies. If the octahedral rotations are condensed, the total energy of the system is significantly reduced and both distortions become dynamically unstable, with $Q_1^{\mathbf{R}}$ slightly favored. This result shows that the approach of a Peierls transition in the $Q_1^{\mathbf{R}}$ coordinate triggered by octahedral rotations \cite{Mercy2017} is equally valid in RMnO$_3$. The argument is point-by-point similar to that in AFeO$_3$ and RNiO$_3$ compounds, as can be found in \cite{Zhang2018,Mercy2017}. We note also that, on the AFM surfaces, we find the same vibronic coupling for $Q_1^{\mathbf{R}}$ as for $Q_{2z}^{\mathbf{M}}$, which we do not show for simplicity. Finally, the competition between $Q_1^{\mathbf{R}}$ as for $Q_{2z}^{\mathbf{M}}$ is decided in favor of $Q_{2z}^{\mathbf{M}}$ by the trilinear coupling with the antiphase rotation $\phi_{xy}^-$ and the antipolar motion $A_X$ (Eq. \eqref{eq:rot_X_Q2}), since there is no such coupling incorporating $Q_1^{\mathbf{R}}$. If the tetragonal and shear strain $Q_{3z}^{\mathbf{\Gamma}}$ and $Q_{4z}^{\mathbf{\Gamma}}$ are relaxed, $Q_1^{\mathbf{R}}$ and $Q_{2z}^{\mathbf{M}}$
get strongly separated (not shown). These results are consistent with the proposed self-trapping of the charge-disproportionated phase \cite{Moskvin2009} and the observation of the coexistence of different phases depending on heat treatments and the history of samples \cite{Huang1997a}. 

\section{Conclusions}   
In conclusion, we presented first-principles calculations able to consistently reproduce the bulk ground state properties of LaMnO$_3$. We systematically investigated the PESs of LaMnO$_3$ around its aristotype cubic reference structure. To do so, we used the decomposition of orthonormal symmetry adapted strains and phonon like modes using \textsc{isodistort}\cite{Campbell2006}. We connected those strains and modes with \textit{Van Vleck's} notation of Jahn-Teller distortion in the isolated octahedral transition-metal complex. We introduced a canonical notation that shows in a simple way the local and cooperative character of such distortions.

The investigation of the $Q_{2z}^{\mathbf{M}}$ PES in the cubic phase by our first principles calculations showed that the origin of the vibronic coupling in the $Q_{2z}^{\mathbf{M}}$ coordinate lies in a Peierls effect. The Peierls condition is met in the AFM-A phase when the spin symmetry is broken along the cubic z-direction. Then, $Q_{2z}^{\mathbf{M}}$ distortion and AFM-A order break all thee translational symmetries and open a band gap. 
 
Through the analysis of the PESs under the presence of other significant lattice distortions that appear in the \textit{Pbnm} phase of LaMnO$_3$, we were able to explain a number of interlocked mechanisms between strain/phonon like distortions, magnetic ordering and the opening of an electronic band gap.
Amongst these, the most important are : 

(i) Octahedral rotations trigger the $Q_{2z}^{\mathbf{M}}$ mode on the FM surface by a negative biquadratic coupling and the AFM surfaces by an increase of the vibronic coupling. The origin of both is the reduced $e_g$ bandwith.

(ii) The most important structural parameter for stabilizing FM over AFM-A magnetic ordering is the tetragonal strain $Q_{3z}^{\mathbf{\Gamma}}$. Reducing this strain will favor the FM state serving as paradigm for engineering FM phases in rare-earth manganites.

(iii) The minimum of FM and AFM-A surfaces have the same structural distortion. This explains the absence of any structural transformations at the AFM to PM transition at $T_N = 140 K$.

Then, we went further and showed from  MC simulations that the orbital ordering transition at $T_{JT} = 750 K$ can be qualitatively reproduced by the PESs provided by our DFT calculations. The analysis of this transition showed mixed characteristics of order-disorder, lattice improper and electronically induced transitions. In this view, the electronic driving force of the MIT can be attributed to \textit{spin symmetry breaking} in the PM metallic phase. It enables the large electron-phonon coupling to explain the high MIT in LaMnO$_3$. Although the spin symmetry breaking appears \textit{dynamically} in the PM phase, its importance lies more in the instantaneous symmetry breaking, which can hence be extracted from \textit{static} DFT calculations, than in its dynamical nature. Nonetheless, this does not rule out any contribution from dynamical electron correlations in stabilising the insulating phase since appropriate U correction to DFT remains important for treating properly LaMnO$_3$.

Finally we showed from first-principles that a subtle competition between charge-ordering and orbital ordering exists in LaMnO$_3$, which further enrich its behavior.
 
While we believe that our work will serve as a sound basis for general lattice-electronic dependencies in LaMnO$_3$ and related compounds, we are aware that not all questions have been addressed. Especially the dynamic nature of \textit{O} phase and the precise mechanism of the orbital ordering transition remain highly debated and we emphasize the need for new general predictive model descriptions. Our work highlights that such model needs to describe self-consistently the \textit{interplay} between lattice and electronic degrees of freedom. A promising tool to achieve such a model description is the generation of so called second-principles model transferring first-principle results into coupled lattice and electronic effective models. Such a second principle approach would then provide access to larger-scale simulations at finite temperature with access to complete local information and allow for combined atomic and electronic fluctuations needed to study the cooperative Jahn-Teller effect in its  dynamical complexity.


\begin{center}
    \textbf{Acknowledgements}
\end{center}
The authors thank Pablo-Garcia Fernandez, Eric Bousquet, and Fabio Ricci for insightful discussions. 
This work was supported by F.R.S.-FNRS project HiT4FiT and PROMOSPAN, ARC project AIMED and M-ERA.NET
project SIOX. Computational resources were provided by the Consortium des Equipements
de Calcul Intensif (CECI), funded by the F.R.S.-FNRS under the Grant No. 2.5020.11 and
the Tier-1 supercomputer of the Fédération Wallonie-Bruxelles funded by the
Walloon Region under the Grant No. 1117545. M.S. and Y.Z. acknowledge financial support from FRIA (Grants No.1.E.070.17. and No. 1.E.122.18.). 

\appendix


\section{Fitting of $\mathbf{Q_{2z}^{\mathbf{M}}}$ PES}
\label{sec:Appendix_fit}

In the following, we discuss briefly the parametrization of the $Q_{2z}^{\mathbf{M}}$ surface in a free energy expansion. To do so, we fitted each of the PES in Fig. \ref{fig4:Cub_Tet_Orth_comp} by a polynomial of the shape 

\begin{equation} 
\mathscr{F} = E_0 + \alpha_{JT}  \abs{Q_{2z}^{\mathbf{M}}} + \alpha Q_{2z}^{\mathbf{M}} + \beta (Q_{2z}^{\mathbf{M}})^2 + \gamma (Q_{2z}^{\mathbf{M}})^4, 
\label{eq:first_polynom_append}
\end{equation}  

where the introduction of the absolute function allows to quantify the vibronic couplings independent of linear asymmetries of the whole PES due to the crystal field. By generation of invariant terms using the \textit{INVARIANTS}\cite{Hatch2003} tool, we defined the following free energy expansion :

\begin{widetext}
\begin{eqnarray}
\mathscr{F}(Q_{2z}^{\mathbf{M}}) =& E_0 +  \alpha_{JT}  \abs{Q_{2z}^{\mathbf{M}}} + \alpha_1[(\phi_{xy}^-)^2\phi_z^+] Q_{2z}^{\mathbf{M}} + \alpha_2 (\phi_{xy}^-A_{X})Q_{2z}^{\mathbf{M}} + \alpha_3 (Q_{3z}^{\mathbf{\Gamma}}\phi_{xy}^{-}A_{X})Q_{2z}^{\mathbf{M}} +  \alpha_4 (Q_{4z}^{\mathbf{\Gamma}}\phi_z)Q_{2z}^{\mathbf{M}} \nonumber \\ & + \alpha_5(Q_{4z}^{\mathbf{\Gamma}}A_{X}\phi_{xy}^-)Q_{2z}^{\mathbf{M}}  +\beta_1 (Q_{2z}^{\mathbf{M}})^2+ \beta_2 \phi^2 (Q_{2z}^{\mathbf{M}})^2 +  \beta_3 A_{X}^2 (Q_{2z}^{\mathbf{M}})^2 + \beta_4Q_{3z}^{\mathbf{\Gamma}}(Q_{2z}^{\mathbf{M}})^2+\beta_5(Q_{3z}^{\mathbf{\Gamma}})^2(Q_{2z}^{\mathbf{M}})^2 \nonumber \\ &  +\beta_6(Q_{4z}^{\mathbf{\Gamma}})^2 (Q_{2z}^{\mathbf{M}})^2  + \gamma (Q_{2z}^{\mathbf{M}})^4,
\label{eq:Invariant}
\end{eqnarray}
\end{widetext}

where we denote coefficients of terms that are of first-order in $Q_{2z}^\mathbf{M}$ with $\alpha$, second with $\beta$, and fourth with $\gamma$. All modes have been normalized such that 1 represents their ground-state amplitude, which can be found in Table \ref{tab:2_all_prop}. Since we are not interested in the fourth-order couplings, we wrote only one fourth-order term and we will not list the variation of its value.
Moreover, we used 
\begin{equation}
\phi = \phi_z^+ = \phi_{xy}^-
\label{eq:def_rot}
\end{equation}
in the $\beta_2$ term, to define a total rotations amplitude $\phi$, as we did not vary the rotations individually. Equation \eqref{eq:def_rot} implies that $\beta_2$ is only valid along a line where the ratio of the amplitudes of the rotations $\phi_z^+$ and $\phi_{xy}^-$ is the same as in the ground-state.
$E_0$, the energy at $Q_{2z}^\mathbf{M} = 0$, is a function of the applied structural distortions. It can be decomposed in the following way 

\begin{widetext}
\begin{equation}
E_0 = E_0^{FM} + E_0^{Q_{3z}^{\mathbf{\Gamma}}} + E_0^{Q_{3z}^{\mathbf{\Gamma}}Q_{4z}^{\mathbf{\Gamma}}}+E_0^{\phi}+E_0^{\phi,Q_{3z}^{\mathbf{\Gamma}}}+E_0^{\phi,Q_{3z}^{\mathbf{\Gamma}},Q_{4z}^{\mathbf{\Gamma}}}+E_0^{\phi,A_{X}}+E_0^{\phi,A_{X},Q_{3z}^{\mathbf{\Gamma}}}+E_0^{\phi,A_{X},Q_{3z}^{\mathbf{\Gamma}},Q_{4z}^{\mathbf{\Gamma}}},
\label{Eq-A4}
\end{equation}
\end{widetext}
 
 where each quantity shows individual energy gains or costs with respect to the cubic AFM-A phase dependent of distortions or magnetic orderings in the superscript. As described in the main text,  the individual strains and distortions were applied with their amplitude in the ground-state of LaMnO$_3$. The values of $E_0$ indicate hence the stability or instability of strains and atomic displacements in the FM and AFM-A phase in the absence of the $Q_{2z}^\mathbf{M}$ distortion. Finally, we also investigated the variation of the electronic instability parameter $\alpha_{JT}$ as a function of the other lattice distortions:

\begin{widetext}
\begin{eqnarray}
\alpha_{JT}&= \alpha_{JT}^0(1+(\lambda_{\phi}+\lambda_{\phi+A_{X}}A_{X})~\phi~+(\lambda_{Q_{3z}^{\mathbf{\Gamma}}}+((\lambda_{Q_{3z}^{\mathbf{\Gamma}}+\phi} + \lambda_{Q_{3z}^{\mathbf{\Gamma}}+\phi+A_{X}}A_{X})~\phi  \nonumber\\ &\dots+(\lambda_{Q_{3z}^{\mathbf{\Gamma}}+Q_{4z}^{\mathbf{\Gamma}}}+(\lambda_{Q_{3z}^{\mathbf{\Gamma}}+Q_{4z}^{\mathbf{\Gamma}}+\phi} +\lambda_{Q_{3z}^{\mathbf{\Gamma}}+Q_{4z}^{\mathbf{\Gamma}}+\phi+A_{X}}~A_{X}~)~\phi~)~Q_{4z}^{\mathbf{\Gamma}}~)~Q_{3z}^{\mathbf{\Gamma}}~),
\end{eqnarray} 
\end{widetext}
 where we assume a linear dependence of the $\alpha_{JT}$ to the other lattice distortions. Further studies would need to clarify the explicit dependence of $\alpha_{JT}$ to the surrounding lattice. As mentioned in the main text, $\alpha_{JT}$ is strictly zero on the FM surface, for which reason only its values for AFM-A ordering has been reported in Table \ref{tab:Constant_fit} below.

\begin{table*}[t]
\renewcommand{\arraystretch}{1.5}
\centering 
	\caption{Table of fitted parameters to reproduce the PES in Fig. \ref{fig4:Cub_Tet_Orth_comp}. Top: zero point energies $E_0$, gathering energy gains or losses of condensing individual modes and strains excluding $Q_{2z}^\mathbf{M}$ distortion. Middle : first- and-second order parameters $\alpha$ and $\beta$ gathering linear and quadratic lattice couplings in $Q_{2z}^\mathbf{M}$. Bottom: electronic parameter $\alpha_{JT}$ gathering the variation of the electronic instability, depending of the condensed lattice modes.}
\label{tab:Constant_fit}
\vspace{0.5em}
\begin{ruledtabular}
\begin{tabularx}{\textwidth}{c c c c c c c c c c c c}
&\multicolumn{11}{c}{Zero Point Energies $E_0$}\\
\cline{2-12} 
\noalign{\vskip 0.5em}  
MO & & $E_0^{FM}$ & $E_0^{Q_{3z}^{\mathbf{\Gamma}}}$ & $E_0^{Q_{3z}^{\mathbf{\Gamma}}Q_{4z}^{\mathbf{\Gamma}}}$ & $E_0^{\phi}$  & $E_0^{\phi,Q_{3z}^{\mathbf{\Gamma}}}$ & $E_0^{\phi,Q_{3z}^{\mathbf{\Gamma}},Q_{4z}^{\mathbf{\Gamma}}}$ & $E_0^{\phi,A_{X}}$ & $E_0^{\phi,A_{X},Q_{3z}^{\mathbf{\Gamma}}}$ & $E_0^{\phi,A_{X},Q_{3z}^{\mathbf{\Gamma}},Q_{4z}^{\mathbf{\Gamma}}}$ &  \\
\hline
FM [eV] & & -0.51 & 0.21 & 0.16 & -0.56 & 0 & 0.16 & -0.40 & -0.15 & -0.27& \\
AFM-A [eV]& & - & 0.05 & 0.15 & -0.63 & 0.05 & 0.18 & -0.39 & -0.12 & -0.29& \\
\hline 
&\multicolumn{11}{c}{First- and Second Order Parameters $\alpha \& \beta$}\\
\cline{2-12} 
MO	& $\alpha_1$ & $\alpha_2$ & $\alpha_3$& $\alpha_4 $&$\alpha_5$& $\beta_1$&$\beta_2$&$\beta_3$&$\beta_4$ &$\beta_5$ &$\beta_6$\\
\hline
FM [eV]	& -0.02 & -0.09 & -0.03 & -0.11 & -0.01 & 0.26 & -0.53 & 0.04 & -0.22 & -0.01 & 0.003 \\
AFM-A [eV]	& -0.01 & -0.10 & -0.02 & -0.11 & -0.02 & 0.29 & -0.04 & 0.02 & 0.08 & -0.01 & -0.20\\	
\hline 
&\multicolumn{11}{c}{Electronic Parameter $\alpha_{JT}$ }\\
\cline{2-12} 	
& & \multicolumn{3}{c}{$\lambda$-Cubic}  & \multicolumn{3}{c}{$\lambda_{Q_{3z}^{\mathbf{\Gamma}}}$} & \multicolumn{3}{c}{$\lambda_{Q_{3z}^{\mathbf{\Gamma}} + Q_{4z}^{\mathbf{\Gamma}}}$}&   \\
& & $\alpha_{JT}^0$ & $+\phi$ & $+\phi + A_{X}$ & 0 & $+\phi$ & $+\phi + A_{X}$ & 0 & $+\phi$ & $+\phi + A_{X}$ & \\
\cline{3-5} \cline{6-8} \cline{9-11}
AFM-A [eV] & & -0.74  & -0.20 & 0.01 & -0.09 & 0.15 & -0.03 & 0.10 & -0.18& 0.04&\\  
\end{tabularx}
\end{ruledtabular} 
\end{table*}


\begin{figure*}
    \includegraphics[width=\textwidth]{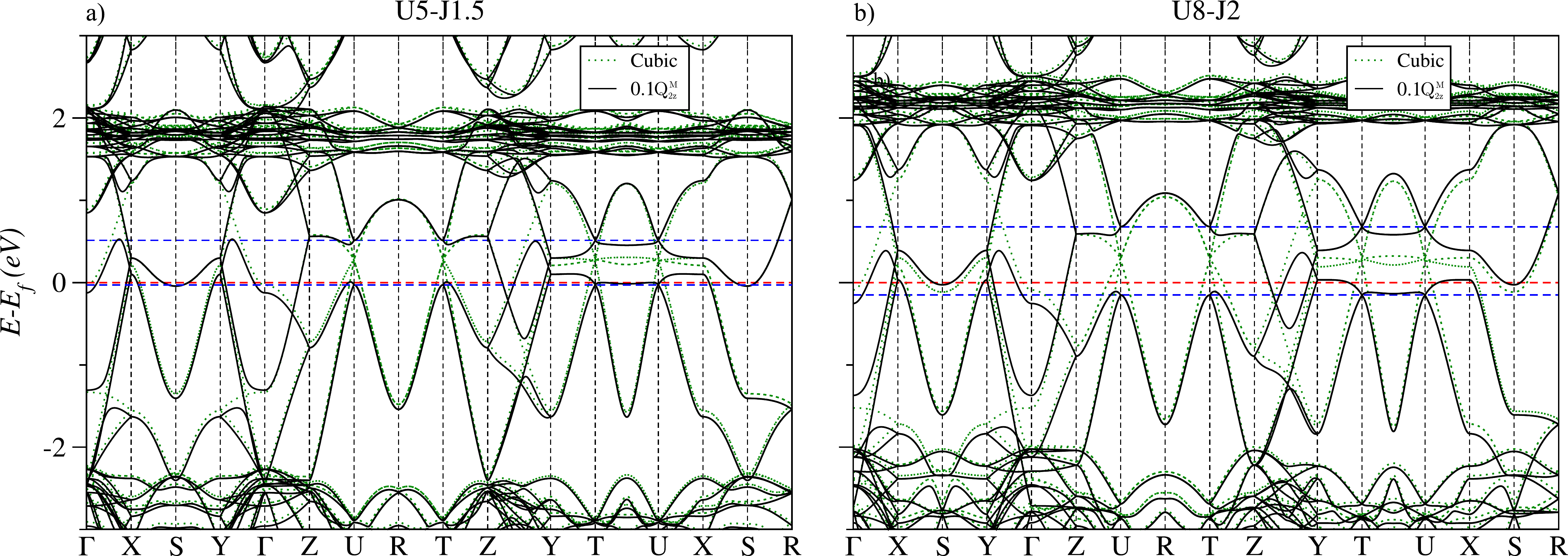}
	\caption{Electronic band structures of the majority spin in FM ordering around the Fermi level. Green dotted lines correspond to the electronic bands in the cubic lattice ($Pm\overline{3}m$) and continuous black lines to the cubic lattice plus 10\% $Q_{2z}^\mathbf{M}$. The red dashed lines shows the Fermi level. Blue dashed lines indicate the band-opening at the high symmetry points $\mathbf{U}$ and $\mathbf{T}$. a) $(U|J)$=(5\si{\electronvolt}$|$1.5\si{\electronvolt}), b)$(U|J)$=(8\si{\electronvolt}$|$2\si{\electronvolt}). }
    \label{fig:UJ_comp_bands}
\end{figure*}
\clearpage 
\pagebreak
\newpage

\section{Effect of (U$|$J) Parameters on the PES}
\label{sec:AppendixII_UJ}

In this section, we provide additional information for the connection between the  $(U|J)$ parameters and the shape of the $Q_{2z}^\mathbf{M}$ PES.
    
Firstly, we explain the different shapes of the PES presented in Fig. \ref{fig3:Exc_comp} in Subsection \ref{subsec:PES_Q2zM}. For this purpose, we show the electronic band-structures of LaMnO$_3$ in its cubic phase (space group $Pm\overline{3}m$) and in a 10\% $Q_{2z}^\mathbf{M}$ distorted structure (space group $P4/mbm$) in FM-ordering using one one hand $(U|J)$ = (5$\si{\electronvolt}|$1.5$\si{\electronvolt}$) and one the other hand $(U|J)$=(8$\si{\electronvolt}|$2$\si{\electronvolt}$) (see Fig. \ref{fig:UJ_comp_bands}). All band structures are presented in the Brillouin-Zone of the 20-atoms supercell, that is the conventional unit-cell for the \textit{Pnma} space-group. The main difference between (5$\si{\electronvolt}|$1.5$\si{\electronvolt}$) and (8$\si{\electronvolt}|$2$\si{\electronvolt}$) in Fig. \ref{fig:UJ_comp_bands} is the splitting energy at gap-openings at high symmetry k-points. These are emphasized with blue dashed lines for the high symmetry points \textbf{U} and \textbf{T}. The splitting energy is significantly increased by the larger $(U|J)$ values (from about 0.5$\si{\electronvolt}$ to 0.8$\si{\electronvolt}$). This difference can explain the different curvature of the FM $Q_{2z}^\mathbf{M}$ PES in Fig. \ref{fig3:Exc_comp} in subsection \ref{subsec:PES_Q2zM}. Through the larger splitting a band gap is opened in a larger part of the Brillouin-Zone for the same $Q_{2z}^\mathbf{M}$ amplitude. The path U-T represents an example. While there is a band just at the Fermi-level in the case of using (5$\si{\electronvolt}|$1.5$\si{\electronvolt}$) which will induce a positve PES curvature, the same band is well below the Fermi level in the case of (8$\si{\electronvolt}|$2$\si{\electronvolt}$) and will therefore contribute to a negative curvature.

\begin{figure}
    \includegraphics[width=0.5\columnwidth]{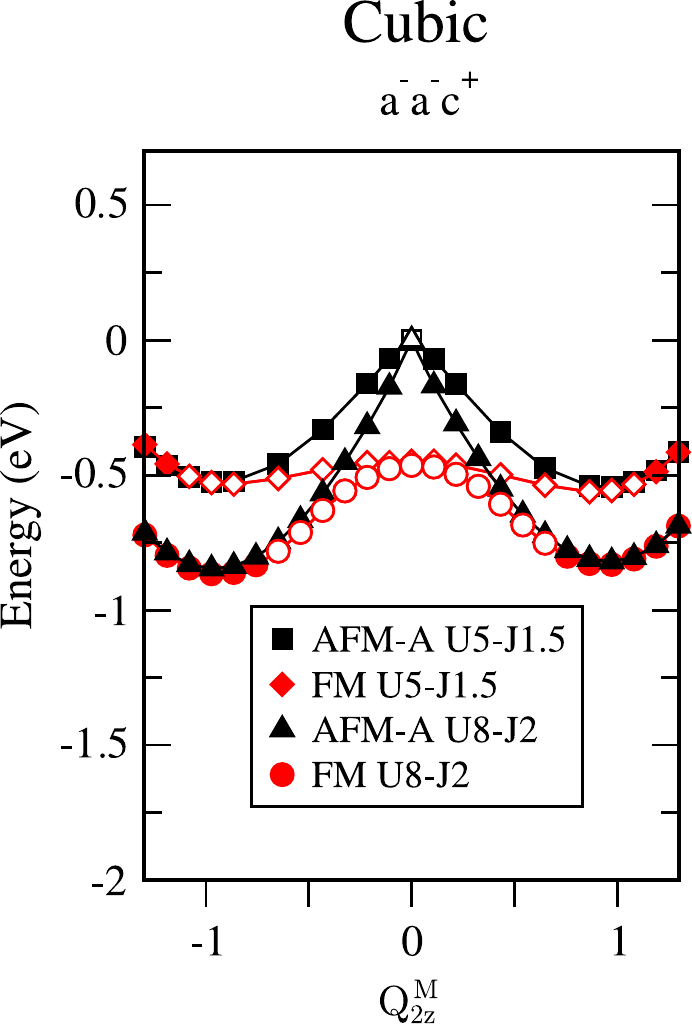}
    \caption{FM and AFM-A PES of $Q_{2z}^\mathbf{M}$ mode within the cubic unit cell and with octahedral rotations condensed with their corresponding ground-state amplitude, for two $(U|J)$ cases. Open (resp. filled) symbols denote metallic (resp. insulating) states.}
    \label{fig:UJ+rot_comp}
\end{figure}

Secondly, we question if changing the $(U|J)$ parameters would \textit{qualitatively} change the results presented in this work. Notably, would a different choice of $(U|J)$ change the results, that the tetragonal strain $Q_{3z}^\mathbf{\Gamma}$ controls the competition between FM and AFM-A? From Fig. \ref{fig3:Exc_comp} in Section \ref{subsec:PES_Q2zM}, we know that in the cubic phase (space group $Pm\overline{3}m$ applying (8$\si{\electronvolt}|$2$\si{\electronvolt}$) keeps the FM ground state.  Here, we additionally show in Fig. \ref{fig:UJ+rot_comp} the $Q_{2z}^\mathbf{M}$ FM and AFM-A PESs within the cubic lattice \textit{with octahedral rotations} condensed with their corresponding ground-state amplitude first using $(U|J)$ = (5$\si{\electronvolt}|$1.5$\si{\electronvolt}$) and then $(U|J)$=(8$\si{\electronvolt}|$2$\si{\electronvolt}$). The first main difference between (5$\si{\electronvolt}|$1.5$\si{\electronvolt}$) and (8$\si{\electronvolt}|$2$\si{\electronvolt}$) is the increased stabilization energy between $Q_{2z}^\mathbf{M}=0$ and the minima positions. The minima positions themselves are not shifted and lie close to the ground-state amplitude of $Q_{2z}^\mathbf{M}$ (=1 in Fig. \ref{fig:UJ+rot_comp}). The second main difference is the insulating electronic ground state around the minima on the FM PES when (8$\si{\electronvolt}|$2$\si{\electronvolt}$) is used. Both these differences can be understood by the analyses of the electronic band structures in dependence of the $(U|J)$ presented in the paragraph above. However, the important common ground between both $(U|J)$ cases are that the FM PES is the lowest in energy. Hence, also in the case of (8$\si{\electronvolt}|$2$\si{\electronvolt}$) octahedral rotations cannot tune the magnetic ground state.  From the analysis in the main text in Section \ref{subsec:PES_rot+AX} and in Appendix \ref{sec:Appendix_fit}, we know that the antipolar motion $A_X$ shows an equal trilinear coupling with $Q_{2z}^\mathbf{M}$ independent of the magnetic order. Therefore it can not tune the competition between AFM-A and FM. This leaves the tetragonal strain $Q_{3z}^\mathbf{\Gamma}$ to be the only structural deformation that can tune competition between AFM-A and FM, which is independent of $(U|J)$.

\clearpage

\bibliographystyle{apsrev4-1}
\bibliography{Bibliography-mend.bib}
\end{document}